\documentstyle[12pt,epsf]{article}
\def\a{\alpha}

\def\p{\phi}

\def\l{\lambda}

\def\th{\theta}

\def\del{\partial}
\def\ha{\frac{1}{2}}
\def\psibar{\overline{\psi}}

\begin{document}
\begin{flushright}
SLAC--PUB--95--7056\\ 
OSU--NT--95--06
\end{flushright}
\bigskip\bigskip
\centerline{{\bf LIGHT-CONE QUANTIZATION AND QCD PHENOMENOLOGY}
            \footnote{\baselineskip=14pt
             Work partially supported by the Department of Energy, contract 
             DE--AC03--76SF00515 and the National Science Foundation
             Grants PHY-9203145, PHY-9258270, and PHY-9207889.}}
\vspace{22pt}
  \centerline{Stanley J. Brodsky}
\vspace{13pt}
  \centerline{\it Stanford Linear Accelerator Center}
  \centerline{\it Stanford University, Stanford, California 94309}
\vspace*{0.5cm}
  \centerline{David G. Robertson}
\vspace{13pt}
  \centerline{\it Department of Physics , The Ohio State University}
  \centerline{\it Columbus, OH 43210}
\vfill
\begin{center}
Submitted to the Proceedings of the\\
``ELFE Summer School and Workshop on Confinement Physics''\\
Cambridge, England \\
22--28 July 1995
\end{center}
\vfill
\pagebreak
\vspace*{0.9cm}
\begin{center} {\bf ABSTRACT} \end{center}
\noindent
In principle, quantum chromodynamics provides a fundamental
description of hadronic and nuclear structure and dynamics in terms
of their elementary quark and gluon degrees of freedom. In practice,
the direct application of QCD to reactions involving the structure
of hadrons is extremely complex because of the interplay of
nonperturbative effects such as color confinement and multi-quark
coherence.  A crucial tool in analyzing such phenomena is the use of
relativistic light-cone quantum mechanics and Fock state methods to
provide tractable and consistent treatments of relativistic
many-body systems.  In this article we present an overview of this
formalism applied to QCD, focusing in particular on applications to
the final states in deep inelastic lepton scattering that will be
relevant for the proposed European Laboratory for Electrons (ELFE),
HERMES, HERA, SLAC, and CEBAF.

We begin with a brief introduction to light-cone field theory,
stressing how it may allow the derivation of a constituent picture,
analogous to the constituent quark model, from QCD.  We then discuss
several applications of the light-cone Fock state formalism to QCD
phenomenology.  The Fock state representation includes all quantum
fluctuations of the hadron wavefunction, including far off-shell
configurations such as intrinsic charm and, in the case of nuclei,
hidden color.  In some applications, such as exclusive processes at
large momentum transfer, one can make first-principle predictions
using factorization theorems which separate the hard perturbative
dynamics from the nonperturbative physics associated with hadron
binding.  The Fock state components of the hadron with small
transverse size, which dominate hard exclusive reactions, have small
color dipole moments and thus diminished hadronic interactions. 
Thus QCD predicts minimal absorptive corrections, i.e., color
transparency for quasi-elastic exclusive reactions in nuclear
targets at large momentum transfer.  In other applications, such as
the calculation of the axial, magnetic, and quadrupole moments of
light nuclei, the QCD relativistic Fock state description provides
new insights which go well beyond the usual assumptions of
traditional hadronic and nuclear physics.
\vfill
\baselineskip 17pt
\pagebreak

\section{QCD on the Light Cone }

One of the central problems in particle physics is to determine the
structure of hadrons such as the proton and neutron in terms of
their fundamental QCD quark and gluon degrees of freedom.  The bound
state structure of hadrons plays a critical role in virtually every
area of particle physics phenomenology.  For example, in the case of
the nucleon form factors, pion electroproduction $ep \rightarrow e
\pi^+n$, and open charm photoproduction $\gamma p\rightarrow
D\Lambda_c$, processes which will be interesting to study at ELFE,
the cross sections depend not only on the nature of the quark
currents, but also on the coupling of the quarks to the initial and
final hadronic states.  Exclusive decay amplitudes such as $B
\rightarrow K^*\gamma$, processes which will be studied intensively
at $B$ factories, depend not only on the underlying weak transitions
between the quark flavors, but also the wavefunctions which describe
how the $B$ and $K^*$ mesons are assembled in terms of their
fundamental quark and gluon constituents.  Unlike the leading twist
structure functions measured in deep inelastic scattering, such
exclusive channels are sensitive to the structure of the hadrons at
the amplitude level and to the coherence between the contributions
of the various quark currents and multi-parton amplitudes.

The analytic problem of describing QCD bound states is compounded
not only by the physics of confinement, but also by the fact that
the wavefunction of a composite of relativistic constituents has to
describe systems of an arbitrary number of quanta with arbitrary
momenta and helicities.  The conventional Fock state expansion based
on equal-time quantization quickly becomes intractable because of
the complexity of the vacuum in a relativistic quantum field theory.
Furthermore, boosting such a wavefunction from the hadron's rest
frame to a moving frame is as complex a problem as solving the bound
state problem itself.  The Bethe-Salpeter bound state formalism,
although manifestly covariant, requires an infinite number of
irreducible kernels to compute the matrix element of the
electromagnetic current even in the limit where one constituent is
heavy.

The description of relativistic composite systems using light-cone
quantization \cite{dirac49} is in contrast remarkably simple.  The
Heisenberg problem for QCD can be written in the form
\begin{equation}
H_{LC }\vert H\rangle = M_H^2 \vert H\rangle\; ,
\end{equation}
where $H_{LC}=P^+ P^- - P_\perp^2$ is the mass operator.  The
operator $P^-=P^0-P^3$ is the generator of translations in the
light-cone time $x^+=x^0+x^3.$ The quantities $P^+=P^0+P^3$ and
$P_\perp$ play the role of the conserved three-momentum.  Each
hadronic eigenstate $\vert H\rangle$ of the QCD light-cone
Hamiltonian can be expanded on the complete set of eigenstates
$\{\vert n\rangle\} $ of the free Hamiltonian which have the same
global quantum numbers: $\vert H\rangle=\sum\psi^H_n(x_i, k_{\perp
i}, \lambda_i) \vert n\rangle.$ In the case of the proton, the Fock
expansion begins with the color singlet state $\vert u u d \rangle $
of free quarks, and continues with $\vert u u d g \rangle $ and the
other quark and gluon states that span the degrees of freedom of the
proton in QCD.  The Fock states $\{\vert n\rangle \}$ are built on
the free vacuum by applying the free light-cone creation operators.
The summation is over all momenta $(x_i, k_{\perp i})$ and
helicities $\lambda_i$ satisfying momentum conservation $\sum^n_i
x_i = 1$ and $\sum^n_i k_{\perp i}=0$ and conservation of the
projection $J^3$ of angular momentum.

The simplicity of the light-cone Fock representation relative to
that in equal-time quantization arises from the fact that the
physical vacuum state has a much simpler structure on the light
cone.  Indeed, kinematical arguments suggest that the light-cone
Fock vacuum is the physical vacuum state.  This means that all
constituents in a physical eigenstate are directly related to that
state, and not disconnected vacuum fluctuations.  In the light-cone
formalism the parton model is literally true.  For example, as we
shall discuss in section 3, all of the structure functions measured
in deep inelastic lepton scattering are simple probabilistic
measures of the light-cone wavefunctions.

The wavefunction $\psi^p_n(x_i, k_{\perp i},\lambda_i)$ describes
the probability amplitude that a proton of momentum $P^+= P^0+P^3$
and transverse momentum $P_\perp$ consists of $n$ quarks and gluons
with helicities $\lambda_i$ and physical momenta $p^+_i= x_i P^+$
and $p_{\perp i} = x_i P_\perp + k_{\perp i}$.  The wavefunctions
$\{\psi^p_n(x_i, k_{\perp i},\lambda_i)\},n=3,\dots$ thus describe
the proton in an arbitrary moving frame.  The variables $(x_i,
k_{\perp i})$ are internal relative momentum coordinates.  The
fractions $x_i = p^+_i/P^+ = (p^0_i+p^3_i)/(P^0+P^3)$, $0 <x_i <1$,
are the boost-invariant light-cone momentum fractions; $y_i= \log
x_i$ is the difference between the rapidity of the constituent $i$
and the rapidity of the parent hadron.  The appearance of relative
coordinates is connected to the simplicity of performing Lorentz
boosts in the light-cone framework.  This is another major advantage
of the light-cone representation.

In principle, the entire spectrum of hadrons and nuclei and their
scattering states is given by the set of eigenstates of the
light-cone Hamiltonian $H_{LC}$ for QCD.  Particle number is
generally not conserved in a relativistic quantum field theory, so
that each eigenstate is represented as a sum over Fock states of
arbitrary particle number. Thus in QCD each hadron is expanded as
second-quantized sums over fluctuations of color-singlet quark and
gluon states of different momenta and number. The coefficients of
these fluctuations are the light-cone wavefunctions $\psi_n(x_i,
k_{\perp i}, \lambda_i).$ The invariant mass ${\cal M}$ of the
partons in a given $n$-particle Fock state can be written in the
elegant form \begin{equation} {\cal M}^2 = \sum_{i=1}^n {k_{\perp
i}^2+m^2\over x_i}\; . \end{equation} The dominant configurations in
the wavefunction are generally those with minimum values of ${\cal
M}^2$.  Note that, except for the case where $m_i=0$ and $k_{\perp
i}=0$, the limit $x_i\rightarrow 0$ is an ultraviolet limit, i.e.,
it corresponds to particles moving with infinite momentum in the
negative $z$ direction: $k^z_i\rightarrow - k^0_i \rightarrow -
\infty.$

The light-cone wavefunctions encode the properties of the hadronic
wavefunctions in terms of their quark and gluon degrees of freedom,
and thus all hadronic properties can be derived from them.  The
natural gauge for light-cone Hamiltonian theories is the light-cone
gauge $A^+=0$.  In this physical gauge the gluons have only two
physical transverse degrees of freedom, and thus it is well matched
to perturbative QCD calculations.

Since QCD is a relativistic quantum field theory, determining the
wavefunction of a hadron is an extraordinarily complex
nonperturbative relativistic many-body problem. In principle it is
possible to compute the light-cone wavefunctions by diagonalizing
the QCD light-cone Hamiltonian on the free Hamiltonian basis.  In
the case of QCD in one space and one time dimensions, the
application of discretized light-cone quantization (DLCQ)
\cite{bp91} provides complete solutions of the theory, including the
entire spectrum of mesons, baryons, and nuclei, and their
wavefunctions \cite{burkardt89,hbp90}.  In the DLCQ method, one
simply diagonalizes the light-cone Hamiltonian for QCD on a
discretized Fock state basis. The DLCQ solutions can be obtained for
arbitrary parameters including the number of flavors and colors and
quark masses.  More recently, DLCQ has been applied to new variants
of QCD$_{1+1}$ with quarks in the adjoint representation, thus
obtaining color-singlet eigenstates analogous to gluonium states
\cite{dkb94}.

The extension of this program to physical theories in 3+1 dimensions
is a formidable computational task because of the much larger number
of degrees of freedom; however, progress is being made.  Analyses of
the spectrum and light-cone wavefunctions of positronium in
QED$_{3+1}$ are given in Ref. \cite{kpw92}.  Currently, Hiller,
Brodsky, and Okamoto \cite{hbo94} are pursuing a nonperturbative
calculation of the lepton anomalous moment in QED using the DLCQ
method. Burkardt has recently solved scalar theories with transverse
dimensions by combining a Monte Carlo lattice method with DLCQ
\cite{burkardt94}.  Also of interest is recent work of Hollenberg
and Witte \cite{Hollenberg}, who have shown how Lanczos
tri-diagonalization can be combined with a plaquette expansion to
obtain an analytic extrapolation of a physical system to infinite
volume.

There has also been considerable work recently focusing on the
truncations required to reduce the space of states to a manageable
level \cite{phw90,perry94,wwhzgp94}.  The natural language for this
discussion is that of the renormalization group, with the goal being
to understand the kinds of effective interactions that occur when
states are removed, either by cutoffs of some kind or by an explicit
Tamm-Dancoff truncation.  Solutions of the resulting effective
Hamiltonians can then be obtained by various means, for example
using DLCQ or basis function techniques.  Some calculations of the
spectrum of heavy quarkonia in this approach have recently been
reported \cite{martina}.

The physical nature of the light-cone Fock representation has
important consequences for the description of hadronic states.  As
we shall discuss in section 3, given the light-cone wavefunctions
$\{\psi_n(x_i, k_{\perp_i},\lambda_i)\}$ one can compute the
electromagnetic and weak form factors from a simple overlap of
light-cone wavefunctions, summed over all Fock states
\cite{dy70,bd80}.  Form factors are generally constructed from
hadronic matrix elements of the current $\langle p \vert j^\mu(0)
\vert p + q\rangle,$ where in the interaction picture we can
identify the fully interacting Heisenberg current $J^\mu$ with the
free current $j_\mu$ at the spacetime point $x^\mu = 0.$

In the case of matrix elements of the current $j^+=j^0+j^3$, in a
frame with $q^+=0,$ only diagonal matrix elements in particle number
$n^\prime = n$ are needed.  In contrast, in the equal-time theory
one must also consider off-diagonal matrix elements and fluctuations
due to particle creation and annihilation in the vacuum.  In the
nonrelativistic limit one can make contact with the usual formulae
for form factors in Schr\"odinger many-body theory.

In the case of inclusive reactions, the hadron and nuclear structure
functions are the probability distributions constructed from
integrals over the absolute squares $\vert \psi_n \vert^2 $, summed
over $n.$ In the far off-shell domain of large parton virtuality,
one can use perturbative QCD to derive the asymptotic fall-off of
the Fock amplitudes, which then in turn leads to the QCD evolution
equations for distribution amplitudes and structure functions. More
generally, one can prove factorization theorems for exclusive and
inclusive reactions which separate the hard and soft momentum
transfer regimes, thus obtaining rigorous predictions for the
leading power behavior contributions to large momentum transfer
cross sections.  One can also compute the far off-shell amplitudes
within the light-cone wavefunctions where heavy quark pairs appear
in the Fock states.  Such states persist over a time $\tau \simeq
P^+/{\cal M}^2$ until they are materialized in the hadron
collisions.  As we shall discuss in section 6, this leads to a
number of novel effects in the hadroproduction of heavy quark
hadronic states \cite{bhmt92}.

Although we are still far from solving QCD explicitly, a number of
properties of the light-cone wavefunctions of the hadrons are known
from both phenomenology and the basic properties of QCD.  For
example, the endpoint behavior of light-cone wavefunctions and
structure functions can be determined from perturbative arguments
and Regge arguments.  Applications are presented in Ref.
\cite{bbs94}.  There are also correspondence principles.  For
example, for heavy quarks in the nonrelativistic limit, the
light-cone formalism reduces to conventional many-body Schr\"odinger
theory.  On the other hand, we can also build effective three-quark
models which encode the static properties of relativistic baryons. 
The properties of such wavefunctions are discussed in section 9.

The remainder of this article is organized as follows.  We begin
with a brief introduction to light-cone quantization, focusing on
its application to solving field theories nonperturbatively.  We
stress the physical nature of the associated Fock space
representation, and discuss how this may allow a connection to be
established between QCD and the constituent quark model.  We then
describe the application of the light-cone formalism to exclusive
processes at large momentum transfer, where factorization theorems
can be used to separate perturbatively calculable hard-scattering
dynamics of the quarks and gluons from the bound-state confinement
dynamics intrinsic to the hadronic wavefunctions.  We briefly touch
on a number of other applications, for example to color
transparency, open charm production, and intrinsic heavy flavors. 
Finally, we discuss the calculation of electromagnetic and weak
moments of nucleons and nuclei in the light-cone framework.

\section{Light-Cone Quantization}

In any practical calculation based on diagonalizing a
field-theoretic Hamiltonian, truncation of the space of states to a
finite subspace is inevitable.  The simplest approach might be to
truncate to the most physically important states, and (numerically)
diagonalize the canonical Hamiltonian on this subspace.  If the
subspace truly contains the states that are most important for
whatever structure is of interest, then the resulting eigenvalues
and wavefunctions should be a reasonably good approximation to the
full solution of the theory. Furthermore, the approximation can be
improved by allowing more and more states into the truncated theory
and verifying that the results converge.

In a more refined approach one would include the effects of the
discarded states in effective interactions.  This step is essential
if one does not have a reliable way of identifying a physically
important subspace {\it a priori}, as in QCD.  It is also very
likely to be the more practical approach.  A useful analogy here
might be with the use of improved actions for lattice gauge theory. 
The lattice spacing $a$ plays the role of an ultraviolet cutoff,
which removes states from the theory with momenta greater than
$\pi/a$.  The problem is that one needs to make $a$ small enough
that low-energy quantities become independent of $a$, but the cost
of a simulation increases rapidly with decreasing $a$, roughly as
$1/a^{\sim(4-7)}$ \cite{lepage9x}. Thus it makes sense to attempt to
remove the dependence on $a$ by modifying the Lagrangian, that is,
by including effective interactions or ``counterterms'' that
incorporate the physics of the states excluded by the cutoff.  This
allows one to work at a larger value of $a$ for a fixed numerical
accuracy, drastically reducing the cost of the simulation.  Of
course, one has to determine the effective interactions to be
included in the Lagrangian.  For QCD this may be done using
perturbation theory if the cutoff is not too low. Asymptotic freedom
implies that the effects of high-energy states are governed by an
effective coupling constant that is small, so that if we eliminate
states of sufficiently high energies then perturbation theory should
suffice.  The resulting perturbatively constructed action can then
be solved nonperturbatively using Monte Carlo techniques.

This kind of Hamiltonian approach is in fact the method of choice in
virtually every area of physics and quantum chemistry.  It has the
desirable feature that the output of such a calculation is
immediately useful: the spectrum of states and wavefunctions. 
Furthermore, it allows the use of intuition developed in the study
of simple quantum systems, and also the application of, e.g.,
powerful variational techniques.  The one area of physics where it
is {\em not} widely employed is relativistic quantum field theory. 
The basic reason for this is that in a relativistic field theory one
has particle creation/annihilation in the vacuum.  Thus the true
ground state is in general extremely complicated, involving a
superposition of states with arbitrary numbers of bare quanta, and
one must understand the complicated structure of this state before
excitations can be considered.  Furthermore, one must have a
nonperturbative way of separating out disconnected contributions to
physical quantities, which are physically irrelevant.  Finally, the
truncations that are required inevitably violate Lorentz covariance
and, for gauge theories, gauge invariance.  It is not clear how to
construct a viable renormalization scheme for this type of problem. 
These difficulties (along with the development of covariant
Lagrangian techniques) eventually led to the almost complete
abandonment of fixed-time Hamiltonian methods in relativistic field
theories.

Light-cone quantization (LCQ) \cite{dirac49} is an alternative to
the usual formulation of field theories in which some of these
problems appear to be more tractable.  This raises the prospect of
developing a practical Hamiltonian approach to solving field
theories, based on diagonalizing LC Hamiltonians.  In the next few
sections we shall give a brief overview of this approach.  We begin
by describing the basic formalism and how it might allow a
connection to be established between QCD and the constituent quark
model.  We then review some existing calculations in toy models, and
finally we discuss the remaining barriers that block progress in
QCD.  Our presentation will necessarily be brief and thus somewhat
superficial.  Our goal is primarily to give a flavor of the LC
approach and why it is of interest, and to set the stage for the
discussion of QCD phenomenology in the following sections.  The
interested reader is advised to consult one of the more extensive
reviews on this subject for detailed discussions of the topics
mentioned here \cite{reviews}.

\subsection{ Basic Formalism }

LCQ is formally similar to equal-time quantization (ETQ) apart from
the choice of initial-value surface.  In ETQ one chooses a surface
of constant time in some Lorentz frame on which to specify initial
values for the fields.  In quantum field theory this corresponds to
specifying commutation relations among the fields at some fixed
time. The equations of motion, or the Heisenberg equations in the
quantum theory, are then used to evolve this initial data in time,
filling out the solution at all spacetime points.

In LCQ one chooses instead a hyperplane tangent to the light
cone---properly called a null plane or light front---as the
initial-value surface.  To be specific we introduce LC coordinates
\begin{equation}
x^\pm \equiv x^0\pm x^3
\end{equation}
(and analogously for all other four-vectors).  The selection of the
3 direction in this definition is of course arbitrary.  Transverse
coordinates will be referred to collectively as $x_\perp = (x^1,
x^2)$.  A null plane is a surface of constant $x^+$ or $x^-.$ It is
conventional to take $x^+$ to be the evolution parameter and choose
as the initial-value surface the null plane $x^+=0$.

In terms of LC coordinates, a contraction of four-vectors decomposes
as
\begin{equation}
p\cdot x = \ha(p^+x^-+p^-x^+)-p_\perp \cdot x_\perp\; ,
\end{equation}
from which we see that the momentum ``conjugate'' to $x^+$ is $p^-$.
Thus the operator $P^-$ plays the role of the Hamiltonian in this
scheme, generating evolution in $x^+$ according to an equation of
the form (in the Heisenberg picture) 
\begin{equation} 
[\p,P^-] = 2i{\del\p\over\del x^+}\; . 
\end{equation}

What is the effect of this new choice of initial-value surface,
apart from the change of coordinates?  The main point is that it
represents a change of {\em representation}, that is, of the Fock
basis used to represent the Hilbert space of a field theory.  The
creation and annihilation operators obtained by projecting fields
onto a null plane create and destroy different states than do the
corresponding operators projected out at equal time.  Furthermore,
the relationship between the LC and ET Fock states is complicated in
an interacting field theory---complicated enough to perhaps be
useful.  A simple way to appreciate this is to imagine starting with
a theory formulated at $t=0$ and solving for the LC Fock states.  To
do this one would evolve the fields to the surface $x^+=0$ and
project out its Fourier modes there.  Because this requires evolving
the fields in time, however, this requires knowing the full solution
of the theory.  Thus the relationship between the two bases is
highly nontrivial, involving the full dynamics of the theory at hand
\cite{rm92}.

There are two main reasons why the LC representation might be useful
in the context of diagonalizing Hamiltonians for quantum field
theories.  First, it can be shown that in LCQ a maximal number of
Poincar\'e generators are kinematic, that is, independent of the
interaction \cite{dirac49,soperphd}.  In ETQ six generators are
kinematical (the momentum and angular momentum operators) and four
are dynamical (the Hamiltonian $H$ and boost generators
${\overrightarrow K}$).  The fact that the boost operators contain
interactions is a serious difficulty, however.  For imagine that we
could actually diagonalize $H$ in some approximation to obtain the
wavefunction for, say, a proton in its rest frame.  Boosting the
state to obtain a moving proton, for use in, e.g., a scattering
calculation, would be quite difficult.  The state transforms as
\begin{equation}
|\psi^\prime\rangle=e^{-i\a K_3}|\psi\rangle
\end{equation}
for a finite boost in the 3-direction, and since $K_3$ is a
complicated operator (as complicated as the Hamiltonian) calculation
of the exponential is difficult.  What is worse is that the
interactions in $K_3$ change particle number.  The boost will
therefore take us out of the truncated space in which we are
working, and a suitable effective boost operator, which acts in the
truncated space, must be constructed.  This may be expected to be as
difficult as that of determining the effective Hamiltonian;
furthermore, there is no reason to expect that the approximations
used to obtain $P^-_{\rm eff}$ will also be appropriate for
constructing the effective boost operator.

As was first shown by Dirac \cite{dirac49}, on the LC seven of the
ten Poincar\'e generators become kinematical, the maximum number
possible. The most important point is that these include Lorentz
boosts.  Thus in the LC representation boosting states is
trivial---the generators are diagonal in the Fock representation so
that computing the necessary exponential is simple.  One result of
this is that the LC theory can be formulated in a manifestly
frame-independent way, yielding wavefunctions that depend only on
momentum fractions and which are valid in any Lorentz frame.  This
advantage is somewhat compensated for, however, in that certain
rotations become nontrivial in LCQ.  Thus rotational invariance will
not be manifest in this approach.

The second advantage of going to the LC is even more striking: the
vacuum state seems to be much simpler in the LC representation than
in ETQ.  Indeed, it is sometimes claimed that the vacuum is
``trivial.'' We shall discuss below to what extent this can really
be true, but for the moment let us give a simple kinematical
argument for the triviality of the vacuum.  We begin by noting that
the longitudinal momentum $p^+$ is conserved in interactions.  For
particles, however, this quantity is strictly positive,
\begin{equation} 
p^+=\left(p_3^2+p_\perp^2+m^2\right)^\ha + p^3 > 0\; . 
\end{equation} 
Thus the Fock vacuum is the only state in the theory with $p^+=0$,
and so it must be an exact eigenstate of the full interacting
Hamiltonian. Stated more dramatically, the Fock vacuum in the LC
representation is the {\em physical} vacuum state.

To the extent that this is really true, it represents a tremendous
simplification, as attempts to compute the spectrum and
wavefunctions of some physical state are not complicated by the need
to recreate a ground state in which processes occur at unrelated
locations and energy scales.  Furthermore, it immediately gives a
constituent picture; all the quanta in a hadron's wavefunction are
directly connected to that hadron.  This allows a precise definition
of the partonic content of hadrons and makes interpretation of the
LC wavefunctions unambiguous.  It also raises the question, however,
of whether LC field theory can be equivalent in all respects to
field theories quantized at equal times, where nonperturbative
effects often lead to nontrivial vacuum structure.  In QCD, for
example, there is an infinity of possible vacua labelled by a
continuous parameter $\theta$, and chiral symmetry is spontaneously
broken.  The question is how it is possible to identify and
incorporate such phenomena into a formalism in which the vacuum
state is apparently simple.

One clue as to how the physics associated with the vacuum can
coexist with a simple vacuum state is provided by the following
series of observations \cite{simplevac,wwhzgp94,reviews}.  In LC
coordinates the free-particle dispersion relation takes the form
\begin{equation}
p^- = {p_\perp^2+m^2\over p^+}\; ,
\label{disprel}
\end{equation}
from which we see that particle states that can combine to give a
complicated vacuum (i.e., that have $p^+\sim0$) are {\em
high-energy} states.\footnote{We ignore for the moment quanta for
which the dispersion relation (\ref{disprel}) does not hold, i.e.,
massless particles with $p_\perp=0$.}  Thus an effective Hamiltonian
approach is natural.  For example, we can introduce an explicit
cutoff on longitudinal momentum for particles:
\begin{equation}
p^+>\l\; .
\end{equation}
This immediately gives a trivial vacuum and the corresponding
constituent picture.  Since the states thus eliminated are
high-energy states, their effects may be incorporated in effective
interactions in the Hamiltonian; the effective interactions they
mediate will be local in (LC) time, so that they can be expressed as
the integral of some Hamiltonian density over the initial-value
surface.\footnote{It is instructive to contrast this with the
situation in ET field theory. Here, many of the states that are
kinematically allowed to mix with the bare vacuum are low-energy
states, so that a description of vacuum physics in terms of
effective Hamiltonians is not practicable.}  In this approach one
can consider the problem of the vacuum as part of the
renormalization problem, that is, the problem of removing dependence
on the cutoff $\l$ from the theory.

Quanta that do not obey Eq. (\ref{disprel}) can simultaneously have
$p^+=0$ and low LC energies, and these may give rise to nontrivial
vacuum structure that cannot be expressed in the form of effective
interactions.  Experience with model field theories, however,
suggests that even in this case the physical vacuum state has a
significantly simpler structure than in ETQ \cite{mccartor91}.  In
addition, these states constitute a set of measure zero in a
(3+1)-dimensional theory. We shall elaborate on this somewhat when
we return to the vacuum problem below.

\subsubsection{Connection to the Constituent Quark Model}

The simplicity of the vacuum means that a powerful physical
intuition can be applied in the study of light-cone QCD: that of the
constituent quark model (CQM).  Indeed, LCQ offers probably the only
realistic hope of deriving a constituent {\em approximation} to QCD,
as stressed particularly by Wilson \cite{wilson,wwhzgp94}.  In
contrast, in equal-time quantization the physical vacuum involves
Fock states with arbitrary numbers of quanta, and a sensible
description of constituent quarks and gluons requires quasi-particle
states, i.e., collective excitations above a complicated ground
state.  Thus an ET approach to hadronic structure based on a few
constituents, analogous to the CQM, is bound to fail.

On the LC, a simple cutoff on small longitudinal momenta suffices to
make the vacuum completely trivial.  Thus we immediately obtain a
constituent picture in which all partons in a hadronic state are
connected directly to the hadron, instead of being disconnected
excitations in a complicated medium.  Whether or not the resulting
theory allows reasonable approximations to hadrons to be constructed
using only a {\it few} constituents is an open question.  However,
one might choose to regard the relative success of the CQM as a
reason for optimism.

The price we pay to achieve this constituent framework is that the
renormalization problem becomes considerably more complicated on the
LC.  We shall discuss this in more detail in section 1.3; for the
moment let us merely note that this is where the familiar ``Law of
Conservation of Difficulty'' manifests itself in the LC approach.

Wilson and collaborators have recently advocated an approach to
solving the light-cone Hamiltonian for QCD which draws heavily on
the physical intuition provided by the CQM \cite{wwhzgp94,perry}. 
One begins by constructing a suitable effective Hamiltonian for QCD,
including the counterterms that remove cutoff dependence.  At
present this can only be done perturbatively, so that the cutoff
Hamiltonian is given as a power series in the coupling constant
$g_\Lambda$:
\begin{equation}
P^-_\Lambda = P^-_{(0)} + g_\Lambda P^-_{(1)}+g_\Lambda^2 P^-
_{(2)}+ \dots\; .
\end{equation}
In the next step a similarity transformation is applied to this
Hamiltonian, which is designed to make it look as much like a CQM
Hamiltonian as possible.  For example, we would seek to eliminate
off-diagonal elements that involve emission and absorption of gluons
or of $q\overline{q}$ pairs.  It is the emission and absorption
processes that are absent from the CQM, so we should remove them by
the unitary transformation.  This procedure cannot be carried out
for all such matrix elements, however.  This is because the
similarity transformation is sufficiently complex that we only know
how to compute it in perturbation theory.  Thus we can reliably
remove in this way only matrix elements that connect states with a
large energy difference; perturbation theory breaks down if we try
to remove, for example, the coupling of a low-energy quark to a
low-energy quark-gluon pair.  We design the transformation
to remove off-diagonal matrix elements between sectors where the
light-cone energy {\it difference} between the initial and final
states is greater than some new cutoff $\lambda$.  This procedure is
known as the ``similarity renormalization group'' method.  For a
more detailed discussion and for connections to RG concepts see
Ref. \cite{gw}.

The result of the similarity transformation is to generate an
effective Hamiltonian $P^-_{\rm eff}$ which has fewer matrix
elements connecting states with different parton number, and
complicated potentials in the diagonal Fock sectors.  The idea is
that the collective states generated in the similarity
transformation will correspond roughly to constituent quarks and
gluons, and the potentials in the different Fock space sectors will
dominate the physics.  If this is correct, then the potentials
should give a reasonable description of hadronic structure, and the
off-diagonal interactions should represent small corrections.  This
can be checked explicitly using bound-state perturbation theory. 
The collective states and potentials would then furnish a
constituent approximation to QCD \cite{perry}.

\subsection{ Applications }

A large number of studies have been performed of model field
theories in the LC framework.  This approach has been remarkably
successful in a range of toy models in 1+1 dimensions: Yukawa theory
\cite{pb85}, the Schwinger model (for both massless and massive
fermions) \cite{schwinger,mccartor91}, $\p^4$ theory \cite{hv8x},
QCD with various types of matter
\cite{burkardt89,hbp90,dkb94,klebanov}, and the sine-Gordon model
\cite{burkardt93}.  It has also been applied with promising results
to theories in 3+1 dimensions, in particular QED \cite{kpw92} and
Yukawa theory \cite{yuk}.  In all cases agreement was found between
the LC calculations and results obtained by more conventional
approaches, for example, lattice gauge theory.  We shall briefly
review two of these applications here: the massless Schwinger model,
and QCD$_{1+1}$ with fundamental fermions.

\subsubsection{ Schwinger Model }

The Schwinger model is simply two-dimensional electrodynamics of
massless fermions.  It is exactly soluble, and the physical spectrum
consists of noninteracting scalar particles.  In addition, the model
possesses a $\th$-vacuum much like that in QCD.  The $\th$-vacuum
breaks chiral symmetry and there is a condensate
\begin{equation}
\langle\th|\psibar\psi|\th\rangle\propto \cos\th\; .
\end{equation}
The presence of nontrivial vacuum structure suggests that the
Schwinger model is a good testing ground for the LC formalism.

In fact all of the known structure of the Schwinger model can be
reproduced in the LC framework \cite{mccartor91}.  There are some
subtleties, however, related to the fact that the LC initial-value
surface is not a good Cauchy surface.  In order to reproduce the
full vacuum structure, fields initialized along a second null plane
(or the equivalent) must be introduced.  In addition, the condensate
one obtains is somewhat sensitive to the precise form of the
infrared regulator, in particular whether or not it breaks parity. 
For a thorough discussion of these issues, see Refs.
\cite{mccartor91,mcc}.

It is interesting to note, however, that if one simply computes the
spectrum of the theory naively in LCQ, without worrying about the
subtleties, then one obtains quite reasonable results
\cite{schwinger}.  Of course this is only possible because the value
of $\th$ has no effect on the spectrum.\footnote{This is a fairly
general feature of two-dimensional models.  There are a number of
theories that possess, e.g., vacuum condensates, but in most of
these the condensate has no physical effect---there is a complete
decoupling of the vacuum and massive sectors \cite{ks95}.}  And
there are many aspects of the model that simply cannot be understood
without addressing the subtleties (the $\th$-vacuum and the anomaly
relation in particular).  Still, it suggests that at least some
quantities may be calculable on the LC without worrying about the
subtleties of the formalism.  It would be very interesting to have a
more general and concrete understanding of this point.

\subsubsection{ QCD$_{1+1}$ with Fundamental Matter }

This theory was originally considered by 't Hooft in the limit of
large $N_c$ \cite{thooft74}.  Later Burkardt \cite{burkardt89}, and
Hornbostel, et al. \cite{hbp90}, gave essentially complete numerical
solutions of the theory for finite $N_c$, obtaining the spectra of
baryons, mesons, and nucleons and their wavefunctions.  The results
are consistent with the few other calculations available for
comparison, and are generally much more efficiently obtained.  In
particular, the mass of the lowest meson agrees to within numerical
accuracy with lattice Hamiltonian results \cite{hamer}.  For $N_c=4$
this mass is close to that obtained by 't Hooft in the
$N_c\rightarrow\infty$ limit \cite{thooft74}.  Finally, the ratio of
baryon to meson mass as a function of $N_c$ agrees with the
strong-coupling results of Ref. \cite{fsxx}.

\begin{figure}
\epsfxsize=3.0in
\centerline{\epsfbox{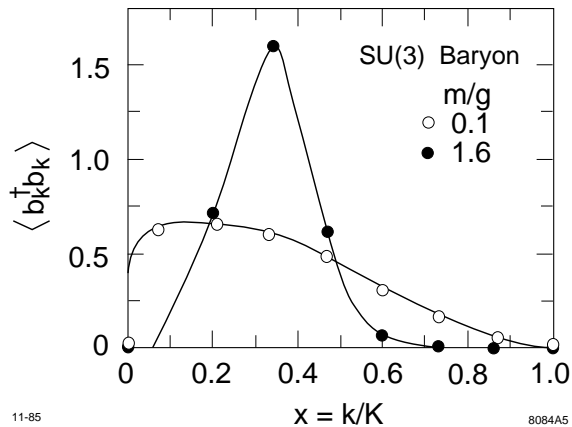}}
\caption{Valence contribution to the baryon structure function in
QCD$_{1+1}$, as a function of the light-cone longitudinal momentum
fraction.  The gauge group is SU(3), $m$ is the quark mass, and $g$
is the gauge coupling. (From Ref. [4].)} 
\label{fig1}
\end{figure}

In addition to the spectrum, of course, one obtains the
wavefunctions. These allow direct computation of, e.g., structure
functions.  (We shall discuss the particularly close relation
between the LC wavefunctions and physical observables in more detail
in the following sections.)  As an example, Fig. 1 shows the valence
contribution to the structure function for an SU(3) baryon, for two
values of the dimensionless coupling $m/g$.  As expected, for weak
coupling the distribution is peaked near $x=1/3$, reflecting that
the baryon momentum is shared essentially equally among its
constituents.  For comparison, the contributions from Fock states
with one and two additional $q\bar{q}$ pairs are shown in Fig. 2. 
Note that the amplitudes for these higher Fock components are quite
small relative to the valence configuration.  The lightest hadrons
are nearly always dominated by the valence Fock state in these
super-renormalizable models; higher Fock wavefunctions are typically
suppressed by factors of 100 or more.  Thus the light-cone quarks
are much more like constituent quarks in these theories than
equal-time quarks would be. As discussed above, in an equal-time
formulation even the vacuum state would be an infinite superposition
of Fock states.  Identifying constituents in this case, three of
which could account for most of the structure of a baryon, would be
quite difficult.

\begin{figure}
\centerline{\epsfbox{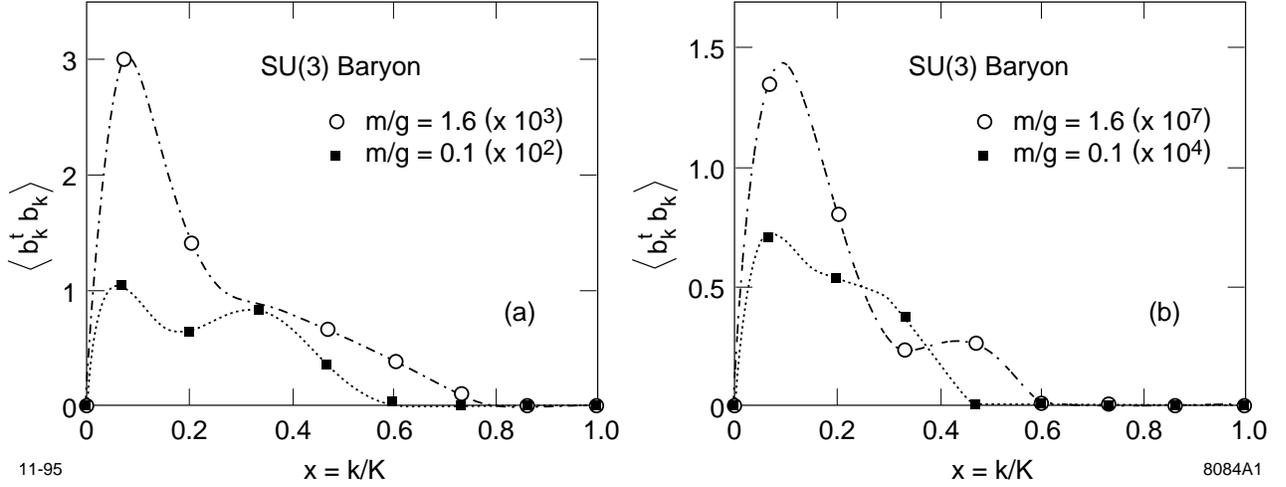}}
\caption{Contributions to the baryon structure function from higher
Fock components: (a) valence plus one additional $q\bar{q}$ pair;
(b) valence plus two additional $q\bar{q}$ pairs. (From Ref. [4].) }
\label{fig2}
\end{figure}

\subsection{Problems and Open Issues }

In this section we briefly survey the main obstacles to progress in
realistic field theories, specifically QCD.  These may be grouped
into three broad categories: the renormalization problem, the
closely related problem of vacuum structure in the LC
representation, and the development of practical algorithms for
calculations in (3+1)-dimensional theories of interest.

\subsubsection{Renormalization }

All nontrivial quantum field theories are afflicted with
divergences, and LC theories are no exception to this.  The theory
must first be made finite by introducing a regulator, and then the
dependence on the unphysical parameters that characterize the
regulator (generically called cutoffs) must be removed by suitably
chosen counterterms. These two problems are of course linked, as the
counterterms required in general depend strongly on the regulator
used.  In particular, it is desirable for a regulator to respect as
many symmetries as possible, so that counterterms will be restricted
to invariant operators.

It is useful to distinguish three generic types of cutoff that are
necessary for LC field theory:\footnote{For actual calculations one
might use a more sophisticated cutoff scheme than that presented
here, for example the ``invariant mass'' cutoff \cite{lb80}, which
preserves the kinematic Lorentz symmetries on the LC, or the
similarity scheme \cite{perrypol}.  The present discussion is merely
intended to highlight the conceptual issues.}
\begin{itemize}
\item A cutoff on light-cone energies: ${p_\perp^2+m^2\over p^+}
<\Lambda$
\item For massless particles, a cutoff on small longitudinal
momenta: $p^+>\lambda$
\item A possible cutoff on particle number: $n<\nu$
\end{itemize}
All of these remove high-energy states on the LC, so that their
counterterms will be local in LC time.\footnote{That the third
cutoff removes high-energy states follows from the positivity of
longitudinal momenta: any state with a large number of particles
must contain some ``wee'' partons, which have high LC energies. 
This is another significant difference between working on the LC and
at equal times, where states with many partons are not necessarily
high-energy states.}

There are two main difficulties that arise in the determination of
these counterterms.  First, all known regulators that are
nonperturbative and applicable to Hamiltonians violate Lorentz and
gauge invariance.  That this will generically be the case can be
seen by noting that some subset of the Lorentz generators are
dynamical, and thus mix states of different particle number.  Any
truncation that limits particle number will in general violate these
symmetries. Gauge invariance will be broken for essentially the same
reason; in QED, for example, the Ward identity relates Green
functions that involve intermediate states with different particle
content.  This means that the counterterms themselves must also
violate these symmetries, so that physical quantities computed from
the full Hamiltonian can be Lorentz- and gauge-invariant.  This is a
major complication, as it drastically increases the number of
possible operators that can occur.

The other main complication follows from the structure of the
dispersion relation on the LC [Eq. (\ref{disprel})].  Transverse UV
divergences ($p_\perp\rightarrow\infty$) can occur for any value of
$p^+$.  This means that counterterms for these divergences in
general involve functions of ratios of all available longitudinal
momenta.  An analogous result holds also for small-$p^+$
divergences---they can occur for any $p_\perp$, and so counterterms
for these will involve functions of transverse momenta.  Thus there
are in general an infinite number of possible counterterms, even if
we restrict consideration to relevant or marginal operators (in the
sense of the renormalization group).  These problems indicate that
renormalization in LC field theory is significantly more complicated
than in other formulations.  It is here that the familiar ``Law of
Conservation of Difficulty'' asserts itself.

The simplest approach to renormalization is to just compute the
counterterms perturbatively.  In analogy with improved lattice
actions, the idea is that asymptotic freedom should make this
sensible if the cutoff is sufficiently high.  This is potentially
correct for states that are removed by the cutoff $\Lambda$; the
perturbative beta function that controls the dependence of the
effective coupling on $\Lambda$ is negative for QCD.  A perturbative
treatment is probably {\em not} sufficient for removing dependence
on $\l,$ however, except perhaps in certain limited domains.  One
sign that the infrared cutoff is fundamentally different from
$\Lambda$ is that longitudinal momentum rescalings are a Lorentz
boost, and so must be an exact symmetry of the theory.  There can be
no beta function associated with longitudinal scale transformations,
unlike rescalings in the transverse directions.  A more physical
point is that all of the vacuum structure is removed by the cutoff
$\l$.  It is very unlikely that the physics associated with the QCD
vacuum can be recreated in the form of counterterms using only
perturbation theory.

For problems where the structure of the vacuum does not play a
central role, however, such a perturbative treatment might be quite
useful. For example, in the study of heavy quarkonia one presumably
does not have to have complete control over the vacuum (e.g., a
nearly massless pion and linear long-range potentials) in order to
obtain reasonable results.

A more ambitious approach to the renormalization problem makes use
of Wilson's formulation of the renormalization group (RG)
\cite{rg,perry94}.  Here one studies sequences of Hamiltonians that
are obtained by iterating a RG transformation which lowers the
various cutoffs.  The idea is to search for those trajectories of
Hamiltonians that can be infinitely long.  A Hamiltonian that lies
on such a trajectory will give results that are equivalent to a
Hamiltonian with infinite cutoffs, that is, results that are
cutoff-independent.  With a perturbative implementation of the RG
this method is equivalent to the first.  It is clearly of interest
to develop nonperturbative realizations of the RG for use in LC
field theories.

\subsubsection{ The Vacuum }

The problem of how to incorporate a nontrivial vacuum in LCQ is
closely related to the renormalization problem; all of the structure
of the vacuum is removed by a small-$p^+$ cutoff, and putting this
physics back is one purpose of the infrared counterterms.  We prefer
to consider it separately, however, because conceptually it is a
much different problem than that of removing dependence on a
transverse momentum cutoff.  The vacuum problem is in fact one
aspect of a whole range of puzzles regarding LC field theory, which
can all be traced to the fact that the LC initial-value surface
contains points that are light-like separated.

Mathematically, the subtleties arise because the LC initial-value
surface is a surface of characteristics \cite{steinhardtetc}.
Physically, they arise because points on the surface can be causally
connected.  Thus one may not be completely free to impose initial
conditions on such a surface, for example.  Furthermore, there is a
danger of missing degrees of freedom; in general, initial conditions
on one characteristic surface are not sufficient to determine a
general solution to the problem \cite{mccartor88,norbert}.  These
difficulties are compounded by the fact that the vacuum lives at a
very singular point in the theory.  Near $p^+=0$ states have
diverging free energies, but the density of states and couplings to
other states are also singular.

One way of addressing these issues is to carefully treat the LC
initial-value problem with an infrared regulator that does not make
the vacuum trivial \cite{rotsurf,vacprob}.  The idea is to formulate
the theory with the vacuum degrees of freedom (sometimes called
``zero modes,'' though this phrase has several distinct meanings
among the experts) present, and then to integrate them out.  This is
essentially the small-$p^+$ part of the renormalization problem
discussed above. The goal is to obtain either an effective
Hamiltonian for use with a trivial vacuum or an explicit description
of the vacuum structure in terms of the LC degrees of freedom.

In the past few years there has been significant progress on
understanding the ways in which vacuum structure can be manifest on
the LC.  A consistent mean-field description of spontaneous symmetry
breaking in the $\phi^4_{1+1}$ theory has been obtained \cite{ssb},
as well as a better understanding of certain topological properties
of gauge theories \cite{topol}.  McCartor's operator solution of the
Schwinger model on the LC is also instructive \cite{mccartor91}.  In
particular the structure of the $\theta$-vacua, while not trivial,
is considerably simpler in the LC representation than in ETQ
\cite{morevac}.

\subsubsection{ Tools for Practical Calculations }

In this section we shall briefly sketch several practical tools that
are being developed for use in LC field theories, and some of the
difficulties associated with them.  These are in many ways
complimentary; they address different practical or theoretical
issues. A judicious combination of them will likely be necessary in
an attack on QCD.

\vskip.1in
\noindent{\underbar{\em Discretized Light-Cone Quantization}}

One approach to small-$k^+$ regularization, which is also convenient
for setting up numerical calculations, is that of ``discretized''
light-cone quantization (DLCQ) \cite{pb85}.  In this method one
imposes boundary conditions on the fields in an interval
\begin{equation}
-L \leq x^- \leq L\; .
\end{equation}
This leads to discrete momenta
\begin{equation}
p^+_n={n\pi\over L}\; ,
\end{equation}
where $n=0,2,4,\dots$ for periodic and $n=1,3,5,\dots$ for
anti-periodic boundary conditions.  The Fock space is thus
denumerable, and after imposing, e.g., transverse momentum cutoffs
the system is completely finite.  It can also be shown that when
periodic boundary conditions are allowed, the mode with $n=0$ (the
``zero mode'') is generally not a dynamical variable, but rather is
a constrained functional of the other, dynamical, modes \cite{czm}.
This is important because it implies that the only state in the
theory with $p^+=0$ is the Fock vacuum.  DLCQ is therefore a
particular way of implementing the infrared cutoff that makes the
vacuum trivial.\footnote{This statement requires some qualification
for gauge theories.  In this case certain zero modes of the gauge
field are in fact dynamical, so that there are particle states with
$p^+=0$ in addition to the Fock vacuum \cite{mccartor91,dzm}.  The
physical vacuum is therefore nontrivial, and this structure must
either be confronted or removed by further {\em ad hoc}
truncations.}  However, one must solve the constraints that
determine the zero modes.

Most of the actual LC calculations done to date have employed DLCQ,
as it gives a particularly clean numerical implementation.  This
method has been extremely successful, particularly in 1+1
dimensions.  A serious difficulty with applying DLCQ to 3+1
dimensional models is the rapid growth of the number of states as
the spacetime dimension is increased.  For example, with a single
particle type and eight momentum states in each of the longitudinal
and transverse directions, there are roughly $6\times 10^{15}$
states.  The resulting Hamiltonian is far too large even to be
stored on a computer, much less diagonalized.

There are several approaches to the problem of large basis size,
some of which involve combining DLCQ with the techniques described
below. For example, one can attempt to explicitly ``integrate out''
some of the states, as in the light-front Tamm-Dancoff approach.  An
interesting implementation of this idea involves formulating the
theory in a small transverse volume, or ``pipe'' \cite{burkardt95}.
The modes with $p_\perp\neq0$ are then high-energy modes (the lowest
nonzero momentum is $\sim1/L_\perp$), and for QCD these can be
integrated out using perturbation theory to obtain an effective
Hamiltonian for the remaining $p_\perp=0$ modes.  Thus the full
theory is reduced to an effective (1+1)-dimensional theory, which is
easily solved using DLCQ.  This is the LC analog of the ET
``femto-universe'' \cite{bj}, particularly as exploited by L\"uscher
and van Baal \cite{lvb}.  The main disadvantage is that the
computation of the effective Hamiltonian is only reliable for small
($<1~{\rm fm}$) transverse volumes.  One might hope to
systematically improve this, for example by gradually allowing
low-transverse-momentum states into the theory.

\vskip.1in
\noindent{\underbar{\em Light-Front Tamm-Dancoff}}

LFTD generically refers to integrating out states (higher Fock
components, for example) to obtain an effective Hamiltonian for some
reasonably-sized subspace \cite{phw90,wwhzgp94}.  The resulting
Hamiltonian can then be solved using DLCQ or some other appropriate
technique.  Of course, it is generally not possible to integrate out
the higher Fock states explicitly.  Instead, one attempts to catalog
the operators allowed by the few symmetries that are respected by
the regulators \cite{wilson,wwhzgp94}.  Generally one restricts
attention to operators that are relevant or marginal in the RG
sense.  One then tries to fix the coefficients of these operators
by, e.g., demanding that symmetries be restored in physical
observables.

The main difficulty with this approach is the large number of
possible relevant and marginal operators on the LC.  As discussed
previously, the regulators we are forced to use violate Lorentz and
gauge invariance (although subsets of Lorentz invariance can be
maintained). Thus the counterterms are not constrained by these
symmetries. Furthermore, power counting in LCQ is complicated by the
presence of separate scales in the longitudinal and transverse
directions.  This leads to the appearance of entire functions of
ratios of momenta in the counterterms.  These are severe
complications, and at present it is not known whether this approach
will yield a predictive theory, that is, one that does not require
the determination of a large number of parameters from fits to data.

As discussed above, one can use perturbation theory to determine the
couplings, although one expects this to be inadequate for
small-$p^+$ quanta.  Ultimately, one would like to address these
issues with a nonperturbative formulation of the renormalization
group.  This is a very challenging problem; there are few examples
of nonperturbative RGs in all of physics.  Alternatively, one can
try to study the effects of the small-$p^+$ quanta directly, and so
uncover the most important operators induced by their elimination.

\vskip.1in
\noindent{\underbar{\em The Transverse Lattice}}

A particularly promising approach to practical calculations involves
combining LCQ with the transverse lattice formulation of Bardeen,
Pearson, and Rabinovici \cite{bprxx}.  Here one discretizes the
transverse dimensions $x_\perp$, but leaves the longitudinal plane
$(x^+,x^-)$ continuous.  One then writes down a LC Hamiltonian in
terms of longitudinal gauge fields and transverse link fields (and
any matter fields), and attempts to solve the resulting theory using
a combination of LC and Monte Carlo techniques \cite{burkardttl}.

This formulation is advantageous for several reasons.  A subset of
gauge invariance (in the transverse directions) can be maintained
explicitly, so that the renormalization problem is perhaps more
tractable.  Furthermore, confinement is manifest for finite lattice
spacing.\footnote{In the transverse directions confinement arises
for the same reasons as in the strong-coupling limit of the
Hamiltonian formulation of lattice gauge theory at equal time
\cite{suss}. Longitudinal confinement is always present on the LC.}
One is therefore already in the ``correct'' phase of the theory, and
the challenge is to show that there is no transition to a
deconfining phase as the continuum limit is approached.  Finally, it
turns out that it is not necessary to diagonalize the entire
Hamiltonian to study the lowest states.  Diagonalization of a small
subset of the Hamiltonian which includes nearest- or
next-to-nearest-neighbor transverse interactions is sufficient
\cite{griffin}.

The main difficulty so far with this approach is a technical one. 
For non-Abelian gauge theories the transverse lattice formulation
reduces to a (1+1)-dimensional gauged nonlinear sigma model (NLSM)
at each transverse site, coupled to their immediate neighbors
\cite{bprxx}. Furthermore, these NLSMs are integrable theories, so
that their exact solutions are in principle known \cite{nlsm}.  The
problem is translating the known solutions into a representation
suitable for application of the LC techniques.  Steps in this
direction have been taken by Griffin \cite{griffin}, but a complete
solution to the problem is lacking.  Given the potential advantages
of the transverse lattice for studies of QCD, we consider this to be
a very important outstanding problem.

\subsection{ The Road to QCD }

The very successful application of the LC formalism to toy models
and QED$_{3+1}$ is encouraging, but much work remains to be done
before a full attack on QCD can begin.  There has recently been
progress on a variety of problems that are important in this regard.
There has been a great deal of work on understanding how vacuum
structure is manifested in principle when quantizing on a null
plane, and how to extract effective Hamiltonians that capture this
structure for use with a simple vacuum state.  There has also been
progress on formulating renormalization groups for LC field theory,
as well as perturbatively constructing LC Hamiltonians for QCD. 
These exhibit some interesting features, for example the natural
appearance of a confining potential \cite{perrypol}, and can be
expected to be useful in, e.g., studies of heavy quarkonia. 
Calculations using these Hamiltonians have recently been reported
\cite{martina}.

Future research will likely proceed along the broad pathways
discussed above.  There is a need for nonperturbative calculations
of the effective operators that occur in the Hamiltonian when the
vacuum structure is eliminated.  This requires either a
nonperturbative RG or a nonperturbative solution to the ``zero
mode'' problem.  In addition, there are challenging technical and
numerical issues that arise in (3+1)-dimensional models---even with
a trivial vacuum, QCD is still an enormously complicated many-body
problem.  The formulation of QCD on a transverse lattice is
particularly relevant in this regard, as it offers a hope of
bringing the explosive growth of the basis size under control.

\section{ Measures of Light-Cone Wavefunctions}

One of the remarkable simplicities of the LC formalism is the fact
that one can write down exact expressions for the spacelike
electromagnetic form factors $\langle P+Q\vert J^+ \vert P\rangle$
of any hadrons for any initial or final state helicity.  At a fixed
light-cone time, the exact Heisenberg current can be identified with
the free current $j^+$.  It is convenient to choose the frame in
which $q^+=0$ so that $q_\perp^2$ is $Q^2 = -q_\mu^2.$ Since the
quark current $j^+$ has simple matrix elements between free Fock
states, each form factor for a given helicity transition
$\lambda\rightarrow \lambda^\prime$ can be evaluated from simple
overlap integrals of the light-cone wavefunctions \cite{dy70,bd80}:
\begin{equation} 
F_{\lambda^\prime, \lambda}(Q^2)= \sum_n\int \prod
d^2 k_{\perp i} \int{\prod d x_i} \overline
\psi_{n,\lambda^\prime}(x_i, k^{\prime}_{\perp i},\lambda_i)
\psi_{n,\lambda}(x_i,k_{\perp i},\lambda_i)\; , 
\end{equation} 
where the integrations are over the unconstrained relative
coordinates. The internal transverse momenta of the final state
wavefunction are $k^{\prime}_\perp = k_\perp + (1-x) q_\perp$ for
the struck quark and $k_\perp^{\prime} = k_\perp -x q_\perp$ for the
spectator quarks.

The structure functions of a hadron can be computed from the square
integral of its LC wavefunctions \cite{lb80}.  For example, the
quark distribution measured in deep inelastic scattering at a given
resolution $Q^2$ is
\begin{equation}
q(x_{Bj},Q^2)= \sum_n \int^{k_\perp^2 <Q^2} \prod d^2 k_{\perp i}
\int{\prod d x_i} \vert\psi_n(x_i, k_{\perp i},\lambda_i)\vert^2
\delta(x_q=x_{Bj})\; ,
\end{equation}
where the struck quark is evaluated with its light-cone fraction
equal to the Bjorken variable: $x_q = x_{Bj}=Q^2/2 p \cdot q.$ A
summation over all contributing Fock states is required to evaluate
the form factors and structure functions.

\section{ Exclusive Processes and Light-Cone Quantization }

A central focus of QCD studies at ELFE will be hadron physics at the
amplitude level.  Exclusive reactions such as pion electroproduction
$\gamma^* p \rightarrow n p$ are more subtle to analyze than deep
inelastic lepton scattering and other leading-twist inclusive
reactions since they require the consideration of coherent QCD
effects.  Nevertheless, there is an extraordinary simplification: In
any exclusive reaction where the hadrons are forced to absorb large
momentum transfer $Q$, one can isolate the nonperturbative
long-distance physics associated with hadron structure from the
short-distance quark-gluon hard scattering amplitudes responsible
for the dynamical reaction.  In essence, to leading order in $1/Q,$
each exclusive reaction $AB \rightarrow CD$ factorizes in the form:
\begin{equation}
T_{AB \rightarrow CD}= \int^1_0 \Pi dx_i \phi^\dagger_D(x_i,Q)
\phi^\dagger_C(x_i,Q) \phi_A(x_i,Q) \phi_B(x_i,Q)
T_{\rm quark}\; ,
\end{equation}
where $\phi_A(x_i,Q) = \int^{k_\perp^2 <Q^2} \prod d^2 k_{\perp i}
{\prod d x_i} \psi_{\rm valence}(x_i, k_{\perp i},\lambda_i)$ is the
process-independent distribution amplitude---the light-cone
wavefunction which describes the coupling of hadron $A$ to its
valence quark with longitudinal light-cone momentum fractions $ 0<
x_i < 1$ at impact separation $b= {\cal O} (1/Q)$---and $T_{\rm
quark}$ is the amplitude describing the hard scattering of the
quarks collinear with the hadrons in the initial state to the quarks
which are collinear with the hadrons in the final state.  Since the
propagators and loop momenta in the hard scattering amplitude
$T_{\rm quark}$ are of order $Q$, it can be computed perturbatively
in QCD.  The dimensional counting rules \cite{bf75} for form factors
and fixed CM scattering angle processes follow from the nominal
power-law falloff of $T_{\rm quark}$. The scattering of the quarks
all occurs at short distances; thus the hard scattering amplitude
only couples to the valence-quarks the hadrons when they are at
small relative impact parameter. Remarkably, there are no initial
state or final state interaction corrections to factorization to
leading order in $1/Q$ because of color coherence; final state color
interactions are suppressed.  This feature not only insures the
validity of the factorization theorem for exclusive processes in
QCD, but it also leads to the novel effect of ``color transparency"
in quasi-elastic nuclear reactions \cite{bbgg81,bm88}.

An essential element of the factorization of high momentum transfer
exclusive reactions is universality, i.e., the distribution
amplitudes $\phi_A(x_i,Q)$ are unique wavefunctions specific to each
hadron. Thus the same wavefunction that controls the meson form
factors also controls the formation of the mesons in exclusive decay
amplitudes of $B$ mesons such as $B \rightarrow \pi \pi$ at the
comparable momenta. The distribution amplitudes obey evolution
equations and renormalization group equations; for details, see Ref.
\cite{lb80}.  A review of the application of light-cone quantized
QCD to exclusive processes is given in Ref. \cite{bl89}.

\section{The Effective Charge $\alpha_V(Q^2)$ and Light-Cone
Quantization } 

The heavy quark potential plays a central role in QCD, not only in
determining the spectrum and wavefunctions of heavy quarkonium, but
also in providing a physical definition of the running coupling for
QCD.  The heavy quark potential $V(Q^2)$ is defined as the
two-particle irreducible amplitude controlling the scattering of two
infinitely heavy test quarks $Q\overline Q$ in an overall
color-singlet state.  Here $Q^2=-q^2={\vec q}^2$ is the momentum
transfer.  The effective charge $\alpha_V(Q^2)$ is then defined
through the relation $V(Q^2) = - 4 \pi C_F\alpha_V(Q^2)/Q^2$ where
$C_F=(N_c^2-1)/2 N_c = 4/3.$ The running coupling $\alpha_V(Q^2)$
satisfies the usual renormalization group equation, where the first
two terms $\beta_0$ and $\beta_1$ in the perturbation series are
universal coefficients independent of the renormalization scheme or
choice of effective charge.  Thus $\alpha_V$ provides a physical
expansion parameter for perturbative expansions in PQCD.

By definition, all quark and gluon vacuum polarization contributions
are summed into $\alpha_V$; the scale $Q$ of $\alpha_V(Q^2)$ that
appears in perturbative expansions is thus fixed by the requirement
that no terms involving the QCD $\beta$-function appear in the
coefficients.  Thus expansions in $\alpha_V$ are identical to that
of conformally invariant QCD.  This argument is the basis for BLM
scale-fixing and commensurate scale relations, which relate physical
observables together without renormalization scale, renormalization
scheme, or other ambiguities arising from theoretical conventions.

There has recently been remarkable progress \cite{davies} in
determining the running coupling $\alpha_V(Q^2)$ from heavy quark
lattice gauge theory using as input a measured level splitting in
the $\Upsilon$ spectrum.  The heavy quark potential can also be
determined in a direct way from experiment by measuring $e^+ e^- \to
c \bar c$ and $e^+ e^- \to b \bar b$ at threshold \cite{bhkt}.  The
cross section at threshold is strongly modified by the QCD
Sommerfeld rescattering of the heavy quarks through their Coulombic
gluon interactions.  The amplitude near threshold is modified by a
factor $S(\beta,Q^2) = x/(1-\exp(-x))$, where
$x=C_F\alpha_V(Q^2)/\beta$ and $\beta=\sqrt{1-4 m_Q^2)/s}$ is the
relative velocity between the produced quark and heavy quark.  The
scale $Q$ reflects the mean exchanged momentum transfer in the
Coulomb rescattering.  For example, the angular distribution for
$e^+ e^- \to Q\overline Q$ has the form $1 + A(\beta)\cos^2
\theta_{\rm cm}.$ The anisotropy predicted in QCD for small $\beta$
is then $A={\widetilde A}/(1+{\widetilde A})$, where
\begin{equation}
 {\widetilde A}=
      {\beta^2\over 2}
      {S(\beta, 4 m_Q^2 \beta^2/e)\over S(\beta, 4 m_Q^2 \beta^2)}
    {1-{4\over \pi} \alpha_V(m_Q^2 \exp 7/6)
         \over 1-{16\over 3 \pi}\alpha_V(m_Q^2 \exp 3/4)}.
\end{equation}
The last factor is due to hard virtual radiative corrections.  The
anisotropy in $e^+ e^- \to Q \overline Q$ will be reflected in the
angular distribution of the heavy mesons produced in the
corresponding exclusive channels.

The renormalization scheme corresponding to the choice of $\alpha_V$
as the coupling is the natural one for analyzing QCD in the
light-cone formalism, since it automatically sums all vacuum
polarization contributions into the coupling.  For example, once one
knows the form of $\alpha_V(Q^2),$ it can be used directly in the
light-cone formalism as a means to compute the wavefunctions and
spectrum of heavy quark systems.  The effects of the light quarks
and higher Fock state gluons that renormalize the coupling are
already contained in $\alpha_V.$

The same coupling can also be used for computing the hard scattering
amplitudes that control large momentum transfer exclusive reactions
and heavy hadron weak decays.  Thus when evaluating $T_{\rm quark}$
the scale appropriate for each appearance of the running coupling
$\alpha_V$ is the momentum transfer of the corresponding exchanged
gluon.  This prescription agrees with the BLM procedure.  The
connection between $\alpha_V$ and the usual $\alpha_{\overline{MS}}$
scheme is described in Ref. \cite{bhjl}.

\section{ The Physics of Light-Cone Fock States }

The light-cone formalism provides the theoretical framework which
allows for a hadron to exist in various Fock configurations.  For
example, quarkonium states not only have valence $Q \overline Q$
components but they also contain $Q\overline Q g$ and $Q \overline Q
g g$ states in which the quark pair is in a color-octet
configuration. Similarly, nuclear LC wave functions contain
components in which the quarks are not in color-singlet nucleon
sub-clusters.  In some processes, such as large momentum transfer
exclusive reactions, only the valence color-singlet Fock state of
the scattering hadrons with small inter-quark impact separation
$b_\perp = {\cal O} (1/Q)$ can couple to the hard scattering
amplitude.  In reactions in which large numbers of particles are
produced, the higher Fock components of the LC wavefunction will be
emphasized. The higher particle number Fock states of a hadron
containing heavy quarks can be diffractively excited, leading to
heavy hadron production in the high momentum fragmentation region of
the projectile. In some cases the projectile's valence quarks can
coalesce with quarks produced in the collision, producing unusual
leading-particle correlations.  Thus the multi-particle nature of
the LC wavefunction can manifest itself in a number of novel ways. 
For example:

\vskip.1in
\noindent\underbar{{\it Color Transparency}}

QCD predicts that the Fock components of a hadron with a small color
dipole moment can pass through nuclear matter without interactions
\cite{bbgg81,bm88}.  Thus in the case of large momentum transfer
reactions, where only small-size valence Fock state configurations
enter the hard scattering amplitude, both the initial and final
state interactions of the hadron states become negligible.  There is
now evidence for QCD ``color transparency" in exclusive virtual
photon $\rho$ production for both nuclear coherent and incoherent
reactions in the E665 experiment at Fermilab \cite{fang93}, as well
as the original measurement at BNL in quasi-elastic $p p$ scattering
in nuclei \cite{heppelmann90}.  In contrast to color transparency,
Fock states with large-scale color configurations interact strongly
and with high particle number production \cite{bbfhs93}.

\vskip.1in
\noindent\underbar{{\it Hidden Color}}

The deuteron form factor at high $Q^2$ is sensitive to wavefunction
configurations where all six quarks overlap within an impact
separation $b_{\perp i} < {\cal O} (1/Q);$ the leading power-law
falloff predicted by QCD is $F_d(Q^2) = f(\alpha_s(Q^2))/(Q^2)^5$,
where, asymptotically, $f(\alpha_s(Q^2))\propto
\alpha_s(Q^2)^{5+2\gamma}$ \cite{bc76}.  The derivation of the
evolution equation for the deuteron distribution amplitude and its
leading anomalous dimension $\gamma$ is given in Ref. \cite{bjl83}.
In general, the six-quark wavefunction of a deuteron is a mixture of
five different color-singlet states. The dominant color
configuration at large distances corresponds to the usual
proton-neutron bound state. However at small impact space
separation, all five Fock color-singlet components eventually
acquire equal weight, i.e., the deuteron wavefunction evolves to
80\%\ ``hidden color.''  The relatively large normalization of the
deuteron form factor observed at large $Q^2$ points to sizable
hidden color contributions \cite{fhzxx}.

\vskip.1in
\noindent\underbar{{\it Spin-Spin Correlations in Nucleon-Nucleon
Scattering and the Charm Threshold}}

One of the most striking anomalies in elastic proton-proton
scattering is the large spin correlation $A_{NN}$ observed at large
angles \cite{krisch92}.  At $\sqrt s \simeq 5 $ GeV, the rate for
scattering with incident proton spins parallel and normal to the
scattering plane is four times larger than that for scattering with
antiparallel polarization. This strong polarization correlation can
be attributed to the onset of charm production in the intermediate
state at this energy \cite{bdt88}.  The intermediate state $\vert u
u d u u d c \bar c \rangle$ has odd intrinsic parity and couples to
the $J=S=1$ initial state, thus strongly enhancing scattering when
the incident projectile and target protons have their spins parallel
and normal to the scattering plane.  The charm threshold can also
explain the anomalous change in color transparency observed at the
same energy in quasi-elastic $ p p$ scattering. A crucial test is
the observation of open charm production near threshold with a cross
section of order of $1 \mu$b.

\vskip.1in
\noindent\underbar{{\it Anomalous Decays of the $J/\psi$}}

The dominant two-body hadronic decay channel of the $J/\psi$ is
$J/\psi \rightarrow \rho \pi$, even though such vector-pseudoscalar
final states are forbidden in leading order by helicity conservation
in perturbative QCD \cite{tuanxx}.  The $\psi^\prime$, on the other
hand, appears to respect PQCD.  The $J/\psi$ anomaly may signal
mixing with vector gluonia or other exotica \cite{tuanxx}.

\vskip.1in
\noindent\underbar{{\it The QCD Van Der Waals Potential and
Nuclear Bound Quarkonium}}

The simplest manifestation of the nuclear force is the interaction
between two heavy quarkonium states, such as the $\Upsilon (b \bar
b)$ and the $J/\psi(c \bar c)$. Since there are no valence quarks in
common, the dominant color-singlet interaction arises simply from
the exchange of two or more gluons. In principle, one could measure
the interactions of such systems by producing pairs of quarkonia in
high energy hadron collisions. The same fundamental QCD van der
Waals potential also dominates the interactions of heavy quarkonia
with ordinary hadrons and nuclei. As shown in Ref. \cite{manoharxx},
the small size of the $Q \overline Q$ bound state relative to the
much larger hadron allows a systematic expansion of the gluonic
potential using the operator product expansion.  The coupling of the
scalar part of the interaction to large-size hadrons is rigorously
normalized to the mass of the state via the trace anomaly. This
scalar attractive potential dominates the interactions at low
relative velocity. In this way one establishes that the nuclear
force between heavy quarkonia and ordinary nuclei is attractive and
sufficiently strong to produce nuclear-bound quarkonium
\cite{manoharxx,btsxx}.

\vskip.1in
\noindent\underbar{{\it Anomalous Quarkonium Production at the
Tevatron}}

Strong discrepancies between conventional QCD predictions and
experiment of a factor of 30 or more have recently been observed for
$\psi$, $\psi^\prime$, and $\Upsilon$ production at large $p_T$ in
high energy $p \overline p$ collisions at the Tevatron \cite{tevxx}.
Braaten and Fleming \cite{bfxx} have suggested that the surplus of
charmonium production is due to the enhanced fragmentation of gluon
jets coupling to the octet $c\overline c$ components in higher Fock
states $\vert c\overline{c}gg\rangle$ of the charmonium
wavefunction. Such Fock states are required for a consistent
treatment of the radiative corrections to the hadronic decay of
$P$-waves in QCD \cite{bblxx}.

\vskip.1in
\noindent\underbar{{\it Intrinsic Heavy Quark Contributions in
Hadron Wavefunctions}}

As we have emphasized, the QCD wavefunction of a hadron can be
represented as a superposition of quark and gluon light-cone Fock
states: $\vert\Psi_{\pi^-}\rangle = \sum_n \psi_{n/\pi^-}
(x_i,k_{\perp i},\lambda_i)\vert n \rangle$, where the color-singlet
states $\vert n \rangle$ represent the Fock components $\vert
\overline u d \rangle$, $\vert \overline u d g \rangle$, $\vert
\overline u d Q \overline Q \rangle$, etc.  Microscopically, the
intrinsic heavy-quark Fock component in the $\pi^-$ wavefunction, $
\vert \overline u d Q \overline Q \rangle$, is generated by virtual
interactions such as $g g \rightarrow Q \overline Q$ where the
gluons couple to two or more projectile valence quarks. The
probability for $Q \overline Q$ fluctuations to exist in a light
hadron thus scales as $\alpha_s^2(m_Q^2)/m_Q^2$ relative to
leading-twist production \cite{vbxx}.  This contribution is
therefore higher twist, and power-law suppressed compared to sea
quark contributions generated by gluon splitting.  When the
projectile scatters in the target, the coherence of the Fock
components is broken and its fluctuations can hadronize, forming new
hadronic systems from the fluctuations \cite{bhmt92}.  For example,
intrinsic $c \overline c$ fluctuations can be liberated provided the
system is probed during the characteristic time $\Delta t = 2p_{\rm
lab}/M^2_{c \overline c}$ that such fluctuations exist. For soft
interactions at momentum scale $\mu$, the intrinsic heavy quark
cross section is suppressed by an additional resolving factor
$\propto \mu^2/m^2_Q$ \cite{chevxx}.  The nuclear dependence arising
from the manifestation of intrinsic charm is expected to be
$\sigma_A\approx \sigma_N A^{2/3}$, characteristic of soft
interactions.

In general, the dominant Fock state configurations are not far off
shell and thus have minimal invariant mass ${\cal M}^2 = \sum_i
m_{T, i}^2/ x_i$ where $m_{T, i}$ is the transverse mass of the
$i^{\rm th}$ particle in the configuration. Intrinsic $Q \overline
Q$ Fock components with minimum invariant mass correspond to
configurations with equal-rapidity constituents. Thus, unlike sea
quarks generated from a single parton, intrinsic heavy quarks tend
to carry a larger fraction of the parent momentum than do the light
quarks \cite{intcxx}.  In fact, if the intrinsic $Q \overline Q$
pair coalesces into a quarkonium state, the momentum of the two
heavy quarks is combined so that the quarkonium state will carry a
significant fraction of the projectile momentum.

There is substantial evidence for the existence of intrinsic $c
\overline c$ fluctuations in the wavefunctions of light hadrons. For
example, the charm structure function of the proton measured by EMC
is significantly larger than that predicted by photon-gluon fusion
at large $x_{Bj}$ \cite{emcicxx}.  Leading charm production in $\pi
N$ and hyperon-$N$ collisions also requires a charm source beyond
leading twist \cite{vbxx,ssnxx}.  The NA3 experiment has also shown
that the single $J/\psi$ cross section at large $x_F$ is greater
than expected from $gg$ and $q \overline q$ production \cite{badxx}.
The nuclear dependence of this forward component is
diffractive-like, as expected from the BHMT mechanism.  In addition,
intrinsic charm may account for the anomalous longitudinal
polarization of the $J/\psi$ at large $x_F$ seen in $\pi N
\rightarrow J/\psi X$ interactions \cite{vanxx}.

Further theoretical work is needed to establish that the data on
direct $J/\psi$ and $\chi_1$ production can indeed be described
using a higher-twist intrinsic charm mechanism, as discussed in Ref.
\cite{bhmt92}.  Experimentally, it is important to check whether the
$J/\psi$'s produced indirectly via $\chi_2$ decay are transversely
polarized.  This would show that $\chi_2$ production is dominantly
leading twist. Better data on real or virtual photoproduction of the
individual charmonium states would also add important information.

\vskip.1in
\noindent\underbar{{\it Double Quarkonium Hadroproduction}}

It is quite rare for two charmonium states to be produced in the
same hadronic collision.  However, the NA3 collaboration has
measured a double $J/\psi$ production rate significantly above
background in multi-muon events with $\pi^-$ beams at laboratory
momentum 150 and 280 GeV/c and a 400 GeV/c proton beam
\cite{badpxx}.  The relative double to single rate, $\sigma_{\psi
\psi}/\sigma_\psi$, is $(3 \pm 1) \times 10^{-4}$ for pion-induced
production, where $\sigma_\psi$ is the integrated single $\psi$
production cross section.  A particularly surprising feature of the
NA3 $\pi^-N\rightarrow\psi\psi X$ events is that the laboratory
fraction of the projectile momentum carried by the $\psi \psi$ pair
is always very large, $x_{\psi \psi} \geq 0.6$ at 150 GeV/c and
$x_{\psi \psi} \geq 0.4$ at 280 GeV/c.  In some events, nearly all
of the projectile momentum is carried by the $\psi \psi$ system! In
contrast, perturbative $ g g$ and $q \overline q$ fusion processes
are expected to produce central $\psi \psi$ pairs, centered around
the mean value, $\langle x_{\psi\psi} \rangle \approx$ 0.4--0.5, in
the laboratory. There have been attempts to explain the NA3 data
within conventional leading-twist QCD. Charmonium pairs can be
produced by a variety of QCD processes including $B \overline B$
production and decay, $B\overline B \rightarrow \psi \psi X$ and
${\cal O}(\alpha_s^4)$ $\psi \psi$ production via $gg$ fusion and $q
\overline q$ annihilation \cite{esxx,russxx}.  Li and Liu have also
considered the possibility that a $2^{++} c\overline c c \overline
c$ resonance is produced, which then decays into correlated
$\psi\psi$ pairs \cite{llxx}.  All of these models predict centrally
produced $\psi \psi$ pairs \cite{bhkxx,russxx}, in contradiction to
the $\pi^-$ data.

Over a sufficiently short time, the pion can contain Fock states of
arbitrary complexity. For example, two intrinsic $c\overline c$
pairs may appear simultaneously in the quantum fluctuations of the
projectile wavefunction and then, freed in an energetic interaction,
coalesce to form a pair of $\psi$'s.  In the simplest analysis, one
assumes the light-cone Fock state wavefunction is approximately
constant up to the energy denominator \cite{vbxx}. The predicted
$\psi \psi$ pair distributions from the intrinsic charm model
provide a natural explanation of the strong forward production of
double $J/\psi$ hadroproduction, and thus gives strong
phenomenological support for the presence of intrinsic heavy quark
states in hadrons.

It is clearly important for the double $J/\psi$ measurements to be
repeated with higher statistics and at higher energies. The same
intrinsic Fock states will also lead to the production of
multi-charmed baryons in the proton fragmentation region. The
intrinsic heavy quark model can also be used to predict the features
of heavier quarkonium hadroproduction, such as $\Upsilon \Upsilon$,
$\Upsilon \psi$, and $(c\bar b)$ $(\bar cb)$ pairs. It is also
interesting to study the correlations of the heavy quarkonium pairs
to search for possible new four-quark bound states and final state
interactions generated by multiple gluon exchange \cite{llxx}, since
the QCD Van der Waals interactions could be anomalously strong at
low relative rapidity \cite{manoharxx,btsxx}.

\vskip.1in
\noindent\underbar{{\it Leading Particle Effect in Open Charm
Production}}

According to PQCD factorization, the fragmentation of a heavy quark
jet is independent of the production process. However, there are
strong correlations between the quantum numbers of $D$ mesons and
the charge of the incident pion beam in $\pi N \rightarrow D X$
reactions. This effect can be explained as being due to the
coalescence of the produced intrinsic charm quark with co-moving
valence quarks. The same higher-twist recombination effect can also
account for the suppression of $J/\psi$ and $\Upsilon$ production in
nuclear collisions in regions of phase space with high particle
density \cite{vbxx}.

There are many ways in which the intrinsic heavy quark content of
light hadrons can be tested. More measurements of the charm and
bottom structure functions at large $x_F$ are needed to confirm the
EMC data \cite{emcicxx}.  Charm production in the proton
fragmentation region in deep inelastic lepton-proton scattering is
sensitive to the hidden charm in the proton wavefunction. The
presence of intrinsic heavy quarks in the hadron wavefunction also
enhances heavy flavor production in hadronic interactions near
threshold.  More generally, the intrinsic heavy quark model leads to
enhanced open and hidden heavy quark production and leading particle
correlations at high $x_F$ in hadron collisions, with a distinctive
strongly shadowed nuclear dependence characteristic of soft hadronic
collisions.

\section{Charm Production at ELFE}

One of the most important areas of experimental investigation at
ELFE will be the production of charm near threshold in
electroproduction and photoproduction, e.g., $\gamma^* p \to J/\psi
p$, $\gamma^* p \to D\Lambda_c$, etc.  These processes are important
to study since they provide new insights into production mechanisms
in QCD and hadronization in a regime where hard gluon radiation is
suppressed. Usually one can rely on the PQCD factorization theorems
for hard exclusive and inclusive processes to accurately compute the
rates for these processes to leading order in $1/m_c$.  In the
low-energy regime accessed by CEBAF and ELFE, however, there can be
significantly modifications to the leading twist QCD predictions:

\begin{itemize}
\item 
The role of intrinsic charm becomes dominant over leading-twist
fusion processes near threshold, since the multi-connected intrinsic
charm configurations in the higher light-cone Fock state of the
proton are more efficient that gluon splitting in producing charm.

\item 
The heavy $c$ and $\bar c$ will be produced at low velocities
relative to each other and with the spectator quarks from the proton
and virtual photon.  As is the case of $e^+ e^- \to \bar c c$ near
threshold, the QCD Coulomb rescattering will give Sommerfeld
correction factors $S(\beta,Q^2)$ which strongly distort the Born
predictions for the production amplitudes. 
\end{itemize}

\section{ Nuclear Effects at ELFE }

The shadowing of the nuclear structure function $F_2^A(x,Q^2)$ at
low $x$ reflects the nuclear dependence of the quark-nucleus cross
section $\sigma_{q n}(\hat s)$ at the corresponding $\hat s = {\cal
O}(\overline{k^2_\perp}/x)$.  Here $\overline{k^2_\perp}$ is the
mean square transverse momentum of the interacting quark.  In the
case of the longitudinal structure function, however, the
leading-twist contribution reflects the interaction of gluons in the
nucleus.  Thus the study of shadowing as a function of photon
polarization can discriminate between the effective quark and gluon
cross sections $\sigma_{g N}(\hat s)$ and $\sigma_{qN}(\hat s),$
fundamental aspects of quark and gluon interactions.  Such a
measurement may be possible at ELFE at high energies.

It is also very interesting to measure the nuclear dependence of
totally diffractive vector meson production $d\sigma/dt(\gamma^*A\to
V A).$ For large photon virtualities (or for heavy vector
quarkonium), the small color dipole moment of the vector system
implies minimal absorption, i.e., color transparency.  Thus,
remarkably, QCD predicts that the forward amplitude $\gamma^* A \to
V A$ at $t \to 0$ is nearly linear in $A$.  One is also sensitive to
corrections from the nonlinear $A$-dependence of the nearly forward
matrix element that couples two gluons to the nucleus, which is
closely related to the nuclear dependence of the gluon structure
function of the nucleus \cite{bgmfs94}.

The integral of the diffractive cross section over the forward peak
is thus predicted to scale approximately as $A^2/R_A^2 \sim
A^{4/3}.$ A test of this prediction could be carried out at very
small $t_{\rm min}$ at HERA, and would provide a striking test of
QCD in exclusive nuclear reactions.  Evidence for color transparency
in quasi-elastic $\rho$ leptoproduction $\gamma^* A \to \rho^0 N
(A-1)$ has recently been reported by the E665 experiment at Fermilab
\cite{e665}.  It is of interest to extend the quasi-elastic
measurements to lower energy at ELFE.

\section{Moments of Nucleons and Nuclei in the Light-Cone Formalism}

The use of covariant kinematics leads to a number of striking
conclusions for the electromagnetic and weak moments of nucleons and
nuclei. For example, magnetic moments cannot be written as the naive
sum $\overrightarrow\mu = \sum\overrightarrow\mu_i$ of the magnetic
moments of the constituents, except in the nonrelativistic limit
where the radius of the bound state is much larger than its Compton
scale: $R_A M_A\gg 1$. The deuteron quadrupole moment is in general
nonzero even if the nucleon-nucleon bound state has no $D$-wave
component \cite{bh83}.  Such effects are due to the fact that even
``static'' moments must be computed as transitions between states of
different momentum $p^\mu$ and $p^\mu + q^\mu$, with $q^\mu
\rightarrow 0$. Thus one must construct current matrix elements
between boosted states. The Wigner boost generates nontrivial
corrections to the current interactions of bound systems
\cite{bp69}.  Remarkably, in the case of the deuteron, both the
quadrupole and magnetic moments become equal to that of the Standard
Model in the limit $M_d R_d\rightarrow 0.$ In this limit, the three
form factors of the deuteron have the same ratios as do those of the
$W$ boson in the Standard Model \cite{bh83}.

One can also use light-cone methods to show that the proton's
magnetic moment $\mu_p$ and its axial-vector coupling $g_A$ have a
relationship independent of the specific form of the light-cone
wavefunction \cite{bs94}.  At the physical value of the proton
radius computed from the slope of the Dirac form factor, $R_1=0.76$
fm, one obtains the experimental values for both $\mu_p$ and $g_A$;
the helicity carried by the valence $u$ and $d$ quarks are each
reduced by a factor $\simeq 0.75$ relative to their nonrelativistic
values. At infinitely small radius $R_p M_p\rightarrow 0$, $\mu_p$
becomes equal to the Dirac moment, as demanded by the
Drell-Hearn-Gerasimov sum rule \cite{gerasimov65,dh66}.  Another
surprising fact is that as $R_1 \rightarrow 0$ the constituent quark
helicities become completely disoriented and $g_A \rightarrow 0$.

One can understand the origins of the above universal features even
in an effective three-quark light-cone Fock description of the
nucleon. In such a model, one assumes that additional degrees of
freedom (including zero modes) can be parameterized through an
effective potential \cite{lb80}.  After truncation, one could in
principle obtain the mass $M$ and light-cone wavefunction of the
three-quark bound-states by solving the Hamiltonian eigenvalue
problem. It is reasonable to assume that adding more quark and
gluonic excitations will only refine this initial approximation
\cite{phw90}. In such a theory the constituent quarks will also
acquire effective masses and form factors.

Since we do not have an explicit representation for the effective
potential in the light-cone Hamiltonian $P^-_{\rm eff}$ for three
quarks, we shall proceed by making an Ansatz for the momentum-space
structure of the wavefunction $\Psi$.  Even without explicit
solutions of the Hamiltonian eigenvalue problem, one knows that the
helicity and flavor structure of the baryon eigenfunctions will
reflect the assumed global SU(6) symmetry and Lorentz invariance of
the theory.  As we will show below, for a given size of the proton
the predictions and interrelations between observables at $Q^2=0,$
such as the proton magnetic moment $\mu_p$ and its axial coupling
$g_A,$ turn out to be essentially independent of the shape of the
wavefunction \cite{bs94}.

The light-cone model given in Ref. \cite{schlumpf93} provides a
framework for representing the general structure of the effective
three-quark wavefunctions for baryons. The wavefunction $\Psi$ is
constructed as the product of a momentum wavefunction, which is
spherically symmetric and invariant under permutations, and a
spin-isospin wave function, which is uniquely determined by
SU(6)-symmetry requirements.  A Wigner-Melosh rotation
\cite{wigner39,melosh74} is applied to the spinors, so that the
wavefunction of the proton is an eigenfunction of $J$ and $J_z$ in
its rest frame \cite{strik,coester82,ls78}.  To represent the range
of uncertainty in the possible form of the momentum wavefunction,
one can choose two simple functions of the invariant mass ${\cal M}$
of the quarks: 
\begin{eqnarray} 
\psi_{\rm H.O.}({\cal M}^2) &=& N_{\rm H.O.}\exp(-{\cal
M}^2/2\beta^2),\\ \psi_{\rm Power}({\cal M}^2) &=& N_{\rm Power}
(1+{\cal M}^2/\beta^2)^{-p}\; ,
\end{eqnarray} 
where $\beta$ sets the characteristic internal momentum scale.
Perturbative QCD predicts a nominal power-law fall off at large
$k_\perp$ corresponding to $p=3.5$ \cite{lb80}.  The Melosh rotation
insures that the nucleon has $j=\ha$ in its rest system.  It has the
matrix representation \cite{melosh74} 
\begin{equation}
R_M(x_i,k_{\perp i},m)={m+x_i {\cal M}-i\overrightarrow
\sigma\cdot(\vec n\times\vec k_i)\over\sqrt{(m+x_i {\cal M})^2+
k_{\perp i}^2} } 
\end{equation} 
with $\vec n=(0,0,1)$, and it becomes the unit matrix if the quarks
are collinear, $R_M(x_i,0,m)=1.$ Thus the internal transverse
momentum dependence of the light-cone wavefunctions also affects its
helicity structure \cite{bp69}.

The Dirac and Pauli form factors $F_1(Q^2)$ and $F_2(Q^2)$ of the
nucleons are given by the spin-conserving and the spin-flip matrix
elements of the vector current $J^+_V$ (at $Q^2=-q^2$) \cite{bd80}
\begin{eqnarray}
F_1(Q^2) &=& \langle p+q,\uparrow | J^+_V |
p,\uparrow \rangle , \\
(Q_1-i Q_2) F_2(Q^2) &=& -2M\langle
p+q,\uparrow | J^+_V | p, \downarrow \rangle \; .
\end{eqnarray}
We then can calculate the anomalous magnetic moment
$a=\lim_{Q^2\rightarrow 0} F_2(Q^2)$.\footnote{The total proton
magnetic moment is $\mu_p = {e \over 2M}(1+a_p).$} The same
parameters as in Ref. \cite{schlumpf93} are chosen, namely $m=0.263$
GeV (0.26 GeV) for the up (down) quark masses, $\beta=0.607$ GeV
(0.55 GeV) for $\psi_{\rm Power}$ ($\psi_{\rm H.O.}$), and $p=3.5$.
The quark currents are taken as elementary currents with Dirac
moments ${e_q \over 2 m_q}.$ All of the baryon moments are well-fit
if one takes the strange quark mass as 0.38 GeV. With the above
values, the proton magnetic moment is 2.81 nuclear magnetons, and
the neutron magnetic moment is $-1.66$ nuclear magnetons. (The
neutron value can be improved by relaxing the assumption of isospin
symmetry.) The radius of the proton is 0.76 fm, i.e., $M_p R_1=3.63$.

In Fig.~3(a) we show the functional relationship between the
anomalous moment $a_p$ and its Dirac radius predicted by the
three-quark light-cone model. The value of
\begin{equation}
R^2_1 = -6 {dF_1(Q^2)\over dQ^2}\Bigl\vert_{Q^2=0}
\end{equation}
is varied by changing $\beta$ in the light-cone wavefunction while
keeping the quark mass $m$ fixed.  The prediction for the power-law
wavefunction $\psi_{\rm Power}$ is given by the broken line; the
continuous line represents $\psi_{\rm H.O.}$.  Figure~3(a) shows
that when one plots the dimensionless observable $a_p$ against the
dimensionless observable $M R_1$ the prediction is essentially
independent of the assumed power-law or Gaussian form of the
three-quark light-cone wavefunction.  Different values of $p>2$ also
do not affect the functional dependence of $a_p(M_p R_1)$ shown in
Fig.~3(a). In this sense the predictions of the three-quark
light-cone model relating the $Q^2 \rightarrow 0$ observables are
essentially model-independent. The only parameter controlling the
relation between the dimensionless observables in the light-cone
three-quark model is $m/M_p$ which is set to 0.28. For the physical
proton radius $M_p R_1=3.63$ one obtains the empirical value for
$a_p=1.79$ (indicated by the dotted lines in Fig. 3(a)).

\setcounter{footnote}{0}

The prediction for the anomalous moment $a$ can be written
analytically as $a=\langle \gamma_V \rangle a^{\rm NR}$, where
$a^{\rm NR}=2M_p/3m$ is the nonrelativistic ($R\rightarrow\infty$)
value and $\gamma_V$ is given as \cite{cc91}
\begin{equation}
\gamma_V(x_i,k_{\perp i},m)=
{3m\over {\cal M}}\left[ {(1-x_3){\cal M}(m+x_3 {\cal
M})- \vec k_{\perp 3}^2/2\over (m+x_3 {\cal M})^2+\vec k_{\perp
3}^2}\right]\; .
\end{equation}
The expectation value $\langle \gamma_V \rangle$ is evaluated
as\footnote{Here $[d^3k]\equiv d\vec k_1d\vec k_2d\vec
k_3\delta(\vec k_1+\vec k_2+ \vec k_3)$. The third component of
$\vec k$ is defined as $k_{3i}\equiv{1\over2}(x_i{\cal M}-{m^2+\vec
k_{\perp i}^2\over x_i {\cal M}})$. This measure differs from the
usual one used in Ref. \cite{lb80} by the Jacobian $\prod
{dk_{3i}\over dx_i}$ which can be absorbed into the wavefunction.}
\begin{equation} 
\langle\gamma_V\rangle = {\int [d^3k] \gamma_V |\psi|^2\over \int
[d^3k] |\psi|^2}\; . 
\end{equation}

Let us now take a closer look at the two limits $R \rightarrow
\infty$ and $R\rightarrow 0$. In the nonrelativistic limit we let
$\beta \rightarrow 0$ and keep the quark mass $m$ and the proton
mass $M_p$ fixed. In this limit the proton radius $R_1 \rightarrow
\infty$ and $a_p \rightarrow 2M_p/3m = 2.38$, since $\langle
\gamma_V \rangle \rightarrow 1$.\footnote{This differs slightly from
the usual nonrelativistic formula $1+a=\sum_q {e_q\over e} {M_p\over
m_q}$ due to the nonvanishing binding energy which results in $M_p
\neq 3m_q$.} Thus the physical value of the anomalous magnetic
moment at the empirical proton radius $M_p R_1=3.63$ is reduced by
25\% from its nonrelativistic value due to relativistic recoil and
nonzero $k_\perp$.\footnote{The nonrelativistic value of the neutron
magnetic moment is reduced by 31\%.}

\begin{figure}
\centerline{\epsfbox{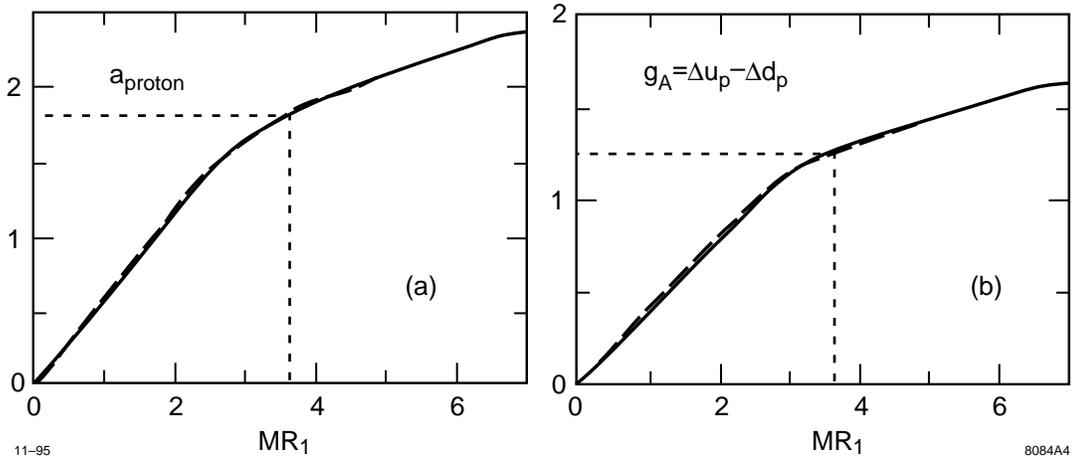}}
\caption{(a).
The anomalous magnetic moment of the proton $a_p=F_2(0)$ as  a
function of its Dirac radius $M_p R_1 $ in Compton units. (b). The
axial vector coupling of the neutron to proton beta-decay as a
function of $M_p R_1.$ In each figure, the broken line is computed
from a wavefunction with power-law falloff  and the solid curve is
computed from a gaussian wavefunction. The experimental values at
the physical proton Dirac radius are indicated by the dotted line.
(From Ref. \ref{bs94}.)}
\label{fig3}
\end{figure}

To obtain the ultra-relativistic limit we let $\beta \rightarrow
\infty$ while keeping $m$ fixed.  In this limit the proton becomes
pointlike, $M_p R_1 \rightarrow 0$, and the internal transverse
momenta $k_\perp \rightarrow\infty$. The anomalous magnetic momentum
of the proton goes linearly to zero as $a=0.43 M_p R_1$ since
$\langle\gamma_V\rangle\rightarrow 0$.  Indeed, the
Drell-Hearn-Gerasimov sum rule \cite{gerasimov65,dh66} demands that
the proton magnetic moment become equal to the Dirac moment at small
radius.  For a spin-${1\over2}$ system
\begin{equation}
a^2={M^2\over 2\pi^2\alpha}\int_{s_{th}}^\infty {ds\over s}\left[
\sigma_P(s)-\sigma_A(s)\right]\; ,
\end{equation}
where $\sigma_{P(A)}$ is the total photoabsorption cross section
with parallel (antiparallel) photon and target spins. If we take the
point-like limit, such that the threshold for inelastic excitation
becomes infinite while the mass of the system is kept finite, the
integral over the photoabsorption cross section vanishes and $a=0$
\cite{bd80}.  In contrast, the anomalous magnetic moment of the
proton does not vanish in the nonrelativistic quark model as
$R\rightarrow 0$. The nonrelativistic quark model does not reflect
the fact that the magnetic moment of a baryon is derived from lepton
scattering at nonzero momentum transfer, i.e., the calculation of a
magnetic moment requires knowledge of the boosted wavefunction.  The
Melosh transformation is also essential for deriving the DHG sum
rule and low-energy theorems of composite systems \cite{bp69}.

A similar analysis can be performed for the axial-vector coupling
measured in neutron decay. The coupling $g_A$ is given by the
spin-conserving axial current $J_A^+$ matrix element
\begin{equation}
g_A(0) =\langle p,\uparrow | J^+_A | p,\uparrow \rangle\; .
\end{equation}
The value for $g_A$ can be written as $g_A=\langle \gamma_A \rangle
g_A^{\rm NR}$, with $g_A^{\rm NR}$ being the nonrelativistic value of
$g_A$ and with $\gamma_A$ given by \cite{cc91,ma91}
\begin{equation}
\gamma_A(x_i,k_{\perp i},m)={(m+x_3 {\cal M})^2-
k_{\perp 3}^2\over (m+x_3 {\cal
M})^2+ k_{\perp 3}^2}\; .
\label{gammaa}
\end{equation}
In Fig.~3(b) the axial-vector coupling is plotted against the proton
radius $M_p R_1$.  The same parameters and the same line
representation as in Fig.~3(a) are used.  The functional dependence
of $g_A(M_p R_1)$ is also found to be independent of the assumed
wavefunction. At the physical proton radius $M_p R_1=3.63$, one
predicts the value $g_A = 1.25$ (indicated by the dotted lines in
Fig.~3(b)), since $\langle \gamma_A \rangle =0.75$.  The measured
value is $g_A= 1.2573\pm 0.0028$ \cite{pdg92}.  This is a 25\%
reduction compared to the nonrelativistic SU(6) value $g_A=5/3,$
which is only valid for a proton with large radius $R_1 \gg 1/M_p.$
As shown in Ref. \cite{ma91}, the Melosh rotation generated by the
internal transverse momentum spoils the usual identification of the
$\gamma^+ \gamma_5$ quark current matrix element with the total
rest-frame spin projection $s_z$, thus resulting in a reduction of
$g_A$.

Thus, given the empirical values for the proton's anomalous moment
$a_p$ and radius $M_p R_1,$ its axial-vector coupling is
automatically fixed at the value $g_A=1.25.$ This is an essentially
model-independent prediction of the three-quark structure of the
proton in QCD.  The Melosh rotation of the light-cone wavefunction
is crucial for reducing the value of the axial coupling from its
nonrelativistic value 5/3 to its empirical value. The near equality
of the ratios $g_A/g_A(R_1 \rightarrow \infty)$ and $a_p/a_p(R_1
\rightarrow \infty)$ as a function of the proton radius $R_1$ shows
the wave-function independence of these quantities.  We emphasize
that at small proton radius the light-cone model predicts not only a
vanishing anomalous moment but also $ \lim_{R_1 \rightarrow 0}
g_A(M_p R_1)=0$.  One can understand this physically: in the zero
radius limit the internal transverse momenta become infinite and the
quark helicities become completely disoriented.  This is in
contradiction with chiral models, which suggest that for a zero
radius composite baryon one should obtain the chiral symmetry result
$g_A=1$.

The helicity measures $\Delta u$ and $\Delta d$ of the nucleon each
experience the same reduction as does $g_A$ due to the Melosh
effect. Indeed, the quantity $\Delta q$ is defined by the axial
current matrix element
\begin{equation}
\Delta q=\langle p,\uparrow | \bar
q\gamma^+\gamma_5 q | p,\uparrow \rangle\; ,
\end{equation}
and the value for $\Delta q$ can be written analytically as $\Delta
q=\langle \gamma_A \rangle \Delta q^{\rm NR}$, with $\Delta q^{\rm
NR}$ being the nonrelativistic or naive value of $\Delta q$ and
$\gamma_A$ given by Eq. (\ref{gammaa}).

The light-cone model also predicts that the quark helicity sum
$\Delta\Sigma=\Delta u+\Delta d$ vanishes as a function of the
proton radius $R_1$. Since $\Delta\Sigma$ depends on the proton
size, it cannot be identified as the vector sum of the rest-frame
constituent spins. As emphasized in Ref. \cite{ma91}, the rest-frame
spin sum is not a Lorentz invariant for a composite system. 
Empirically, one can measure $\Delta q$ from the first moment of the
leading-twist polarized structure function $g_1(x,Q).$ In the
light-cone and parton model descriptions, $\Delta q=\int_0^1 dx
[q^\uparrow (x) - q^\downarrow (x)]$, where $q^\uparrow (x)$ and
$q^\downarrow (x)$ can be interpreted as the probability for finding
a quark or antiquark with longitudinal momentum fraction $x$ and
polarization parallel or antiparallel to the proton helicity in the
proton's infinite momentum frame \cite{lb80}.  [In the infinite
momentum frame there is no distinction between the quark helicity
and its spin projection $s_z.$] Thus $\Delta q$ refers to the
difference of helicities at fixed light-cone time or at infinite
momentum; it cannot be identified with
$q(s_z=+{1\over2})-q(s_z=-{1\over2}),$ the spin carried by each
quark flavor in the proton rest frame in the equal-time formalism.

Thus the usual SU(6) values $\Delta u^{\rm NR}=4/3$ and $\Delta
d^{\rm NR}=-1/3$ are only valid predictions for the proton at large
$M R_1.$ At the physical radius the quark helicities are reduced by
the same ratio 0.75 as is $g_A/g_A^{\rm NR}$ due to the Melosh
rotation. Qualitative arguments for such a reduction have been given
in Refs. \cite{karl92,fritzsch90}.  For $M_p R_1 = 3.63,$ the
three-quark model predicts $\Delta u=1,$ $\Delta d=-1/4,$ and
$\Delta\Sigma=\Delta u+\Delta d = 0.75$.  Although the gluon
contribution $\Delta G=0$ in our model, the general sum rule
\cite{jm90}
\begin{equation}
{1\over2}\Delta\Sigma +\Delta G+L_z= {1\over2}
\end{equation}
is still satisfied, since the Melosh transformation effectively
contributes to $L_z$.

Suppose one adds polarized gluons to the three-quark light-cone
model. Then the flavor-singlet quark-loop radiative corrections to
the gluon propagator will give an anomalous contribution $\delta
(\Delta q)=-{\alpha_s\over2\pi}\Delta G$ to each light quark
helicity \cite{Altarelli}.  The predicted value of $g_A = \Delta
u-\Delta d$ is of course unchanged. For illustration we shall choose
${\alpha_s\over2\pi}\Delta G=0.15$. The gluon-enhanced quark model
then gives the values in Table~1, which agree well with the present
experimental values. Note that the gluon anomaly contribution to
$\Delta s$ has probably been overestimated here due to the large
strange quark mass. One could also envision other sources for this
shift of $\Delta q$ such as intrinsic flavor \cite{fritzsch90}.  A
specific model for the gluon helicity distribution in the nucleon
bound state is given in Ref. \cite{bbs94}.

\begin{table}
\begin{center}
\begin{tabular}{|c|c|c|c|c|}
\hline Quantity & NR & $3q$ & $3q+g$ & Experiment \\ \hline $\Delta u$
& ${4\over3}$ & 1 & 0.85 & $0.83\pm 0.03 $ \\ $\Delta d$ &$-{1\over3}$
& $-{1\over4}$ & --0.40 & $-0.43\pm 0.03 $\\ $\Delta s$ & 0 & 0 &
--0.15 & $-0.10\pm 0.03 $\\ $\Delta \Sigma$ &1 & ${3\over4}$ & 0.30 &
$0.31\pm 0.07 $\\ \hline
\end{tabular}
\end{center}
\caption{Comparison of the quark content of the proton
in the nonrelativistic quark model (NR), in the three-quark model
($3q$), in a gluon-enhanced three-quark model ($3q+g$), and with
experiment [112]. 
}
\end{table}

In the above analysis of the singlet moments, it is assumed that all
contributions to the sea quark moments derive from the gluon anomaly
contribution $\delta (\Delta q)=-{\alpha_s\over2\pi}\Delta G$. In
this case the strange and anti-strange quark distributions will be
identical. On the other hand, if the strange quarks derive from the
intrinsic structure of the proton, then one would not expect this
symmetry.  For example, in the intrinsic strangeness wavefunctions,
the dominant fluctuations in the nucleon wavefunction are most
likely dual to intermediate $\Lambda$-$K$ configurations since they
have the lowest off-shell light-cone energy and invariant mass. In
this case $s(x)$ and $\bar s(x)$ will be different.

The light-cone formalism also has interesting consequences for spin
correlations in jet fragmentation.  In LEP or SLC one produces $s$
and $\bar s$ quarks with opposite helicity.  This produces a
correlation of the spins of the $\Lambda$ and $\overline\Lambda$,
each produced with large $z$ in the fragmentation of their
respective jet. The $\Lambda$ spin essentially follows the spin of
the strange quark since the $ud$ has $J=0$.  However, this cannot be
a 100\% correlation since the $\Lambda$ generally is produced with
some transverse momentum relative to the $s$ jet.  In fact, from the
light-cone analysis of the proton spin, we would expect no more than
a 75\% correlation since the $\Lambda$ and proton radius should be
almost the same. On the other hand if $z=E_\Lambda/E_s \to 1,$ there
can be no wasted energy in transverse momentum.  At this point one
could have 100\% polarization. In fact, the nonvalence Fock states
will be suppressed at the extreme kinematics, so there is even more
reason to expect complete helicity correlation in the endpoint
region.

We can also apply a similar idea to the study of the fragmentation
of strange quarks to $\Lambda$s produced in deep inelastic lepton
scattering on a proton at ELFE.  One could use the correlation
between the spin of the target proton and the spin of the $\Lambda$
to directly measure the strange polarization $\Delta s.$ It is
conceivable that any differences between $\Delta s$ and $\Delta \bar
s$ in the nucleon wavefunction could be distinguished by measuring
the correlations between the target polarization and the $\Lambda$
and $\overline\Lambda$ polarization in deep inelastic lepton proton
collisions at ELFE or in the target polarization region in
hadron-proton collisions.

In summary, we have shown that relativistic effects are crucial for
understanding the spin structure of nucleons. By plotting
dimensionless observables against dimensionless observables, we
obtain relations that are independent of the momentum-space form of
the three-quark light-cone wavefunctions. For example, the value of
$g_A \simeq 1.25$ is correctly predicted from the empirical value of
the proton's anomalous moment. For the physical proton radius $M_p
R_1= 3.63$, the inclusion of the Wigner-Melosh rotation due to the
finite relative transverse momenta of the three quarks results in a
$\sim 25\%$ reduction of the nonrelativistic predictions for the
anomalous magnetic moment, the axial vector coupling, and the quark
helicity content of the proton.  At zero radius, the quark
helicities become completely disoriented because of the large
internal momenta, resulting in the vanishing of $g_A$ and the total
quark helicity $\Delta \Sigma.$

\section{ Future Directions }

The light-cone formalism is a very promising framework for the study
of hadronic structure.  The fact that it allows a precise definition
of the parton model means that the light-cone wavefunctions are the
most natural way of encoding hadronic structure.  The ability to
boost states easily---manifested in the frame-independence of the
formalism---is another major advantage.  Finally, light-cone
quantization offers the best hope for deriving a constituent {\em
approximation} to hadronic structure from QCD.  In any other frame,
the need to understand the constituents as quasi-particles makes
building a connection to a CQM essentially hopeless.

As we have emphasized in these lectures, the proton is represented
in QCD at a given light-cone time $x^+ = t+z$ as a superposition of
quark and gluon Fock states $\vert uud\rangle,\vert uudg\rangle,
\vert uud Q \overline Q\rangle$, etc.  Thus when the proton is
expanded on a free quark and gluon basis, it is a fluctuating system
of arbitrarily large particle number.  The light-cone wavefunctions
$\psi_n(x_i, k_{\perp i}, \lambda_i)$ are the probability amplitudes
which describe the projections of the proton state on this Hilbert
space.  The structure functions measured in deep inelastic lepton
scattering are directly related to the light-cone $x$ momentum
distributions of the quarks and gluons determined by the
$\vert\psi_n\vert^2$.  Another interesting measure of the proton's
structure involves examining the system of hadrons produced in the
proton's fragmentation region when one quark is removed, i.e., the
proton's ``fracture functions'' \cite{fract}. At HERA, the particles
derived from the spectator $\bar 3_C$ system which are intrinsic to
the proton's structure are produced in the proton beam direction
with approximately the same rapidity as that of the proton at
relatively small transverse momentum \cite{ssnxx}.  Thus in
high-energy $e p$ collisions, the electron resolves the
diffractively-excited proton, revealing the correlations of the
spectator quarks and gluons in its light-cone Fock components with
invariant mass extending up to the energy of the collision.

It is of particular interest to examine the fragmentation of the
proton when the electron strikes a light quark and the interacting
Fock component is the $\vert uud c \bar c \rangle$ or $\vert uud b
\bar b \rangle$ state.  These Fock components correspond to
intrinsic charm or intrinsic bottom quarks in the proton
wavefunction.  Since the heavy quarks in the proton bound state have
roughly the same rapidity as the proton itself, the intrinsic heavy
quarks will appear at large $x_F$.  One expects heavy quarkonium and
also heavy hadrons to be formed from the coalescence of the heavy
quark with the valence $u$ and $d$ quarks, since they have nearly
the same rapidity.  Since the heavy and valence quark momenta
combine, these states are preferentially produced with large
longitudinal momentum fractions.

A recent analysis by Harris, Smith and Vogt \cite{hsv95} of the
excessively large charm structure function of the proton at large
$x$ as measured by the EMC collaboration at CERN yields an estimate
that the probability $P_{c \bar c}$ that the proton contains
intrinsic charm Fock states is of the order of 0.6\%.  In the case
of intrinsic bottom, PQCD scaling predicts
\begin{equation}
P_{ b \bar b}=P_{c \bar c} {m^2_\psi\over m^2_\Upsilon}
{\alpha^4_s(m_b) \over \alpha^4_s(m_c)}\; ,
\end{equation}
more than an order of magnitude smaller.  If super-partners of the
quarks or gluons exist they must also appear in higher Fock states
of the proton, such as $\vert uud ~{\rm gluino}~ {\rm
gluino}\rangle$. At sufficiently high energies, the diffractive
excitation of the proton will produce these intrinsic quarks and
gluinos in the proton fragmentation region.  Such supersymmetric
particles can bind with the valence quarks to produce highly unusual
color-singlet hybrid supersymmetric states such as $\vert uud ~{\rm
gluino}\rangle$ at high $x_F.$ The probability that the proton
contains intrinsic gluinos or squarks scales with the appropriate
color factor and inversely with the heavy particle mass squared
relative to the intrinsic charm and bottom probabilities.  This
probability is directly reflected in the production rate when the
hadron is probed at a hard scale $Q$ which is large compared to the
virtual mass ${\cal M}$ of the Fock state.  At low virtualities, the
rate is suppressed by an extra factor of $Q^2/{\cal M}^2.$ The
forward proton fragmentation regime is a challenge to instrument at
HERA, but it may be feasible to tag special channels involving
neutral hadrons or muons.  In the case of the gas jet fixed-target
$ep$ collisions at ELFE or HERMES, the target fragments emerge at
low velocity and large backward angles, and thus may be accessible
to precise measurement.

As we have outlined in these lectures, the light-cone Fock
representation of quantum chromodynamics provides both a tool and a
language for representing hadrons as fluctuating composites of
fundamental quark and gluon degrees of freedom.  Light-cone
quantization provides an attractive method to compute this structure
from first principles in QCD. However, much more progress in theory
and in experiment will be needed to fulfill this promise.

\vspace{.5in}
\centerline{\bf Acknowledgements}

\noindent
It is a pleasure to thank the organizers of the ELFE Workshop and
Summer School, in particular S. D. Bass, for their efforts and
hospitality.  S.J.B. also wishes to thank J. Hiller, P. Hoyer, O.
Jacob, G. P. Lepage, H. J. Lu, A. Mueller, H.-C. Pauli, S. Pinsky,
W.-K. Tang, F. Schlumpf, and R. Vogt for helpful conversations.
D.G.R. is grateful to M. Burkardt, K. Hornbostel and R. J. Perry for
many valuable discussions and comments on portions of the
manuscript. The work of S.J.B. was supported by the Department of
Energy, contract No. DE-AC03-76SF00515.  D.G.R was supported by the
National Science Foundation under Grants Nos. PHY-9203145,
PHY-9258270, and PHY-9207889.

\end{document}